# Destructive interference of second harmonic generation in AA stacked $MoTe_2/WSe_2$

Yiduo Wang[1, #], Yao Lu[2, #], Changshen Chen[1, 3, #], Xiaotong Liao[4, #], Siyu Fan[1, 7], Zhenyu Wang[1, 7], Yaotian Liu[1, 7], Subi Du[1, 7], Yingze Jia[1, 7], Ye Zhu[3], Yingwei Wang[5], Jun He[5], Song Liu[6], Jiawei Ruan[4, *], Zhen Chen[1, 7, *], Kai-Qiang Lin[2, *], Yang Xu[1, 7, *]

[1]Beijing National Laboratory for Condensed Matter Physics, Institute of Physics, Chinese Academy of Sciences, Beijing, 100190, China.

[2]State Key Laboratory of Physical Chemistry of Solid Surfaces, College of Chemistry and Chemical Engineering, Xiamen University, Xiamen, China.

[3]Department of Applied Physics, Research Institute for Smart Energy, The Hong Kong Polytechnic University, Kowloon, Hong Kong, 999077, China.

[4]Eastern Institute of Technology, Ningbo 315200, China

[5]Hunan Key Laboratory of Nanophotonics and Devices, School of Physics and Electronics, Central South University, Changsha 410083, China.

[6]Institute of Microelectronics, Chinese Academy of Science, Beijing 100029, China

[7]School of Physical Sciences, University of Chinese Academy of Sciences, Beijing, China.

#These authors contributed equally to this work.

*Email: jwruan@eitech.edu.cn, zhen.chen@iphy.ac.cn, kaiqiang.lin@xmu.edu.cn, yang.xu@iphy.ac.cn

**ABSTRACT:** The stacking configuration of two-dimensional materials critically governs their optical and electronic responses. Monolayer transition-metal dichalcogenides (TMDC) lack inversion symmetry and exhibit exciton-enhanced second-harmonic generation (SHG). In TMDC bilayers, 60° (0°) stacking is conventionally expected to suppress (enhance) SHG owing to destructive (constructive) interference of the layer-resolved nonlinear polarizations. Here, we report an unconventional destructive SHG interference in nearly 0°-stacked (AA-stacked) $MoTe_2/WSe_2$ heterobilayers using two independent probes: atomic-resolution imaging and stacking-sensitive exciton hybridization measurements. Supported by *ab initio* GW and Bethe-Salpeter equation calculations, we show that distinct two-photon resonances associated with the $WSe_2$ C exciton and the $MoTe_2$ D exciton generate a nearly $\pi$ phase difference ($\Delta\phi$) in their second-order nonlinear susceptibilities $\chi^{(2)}$, leading to the anomalous destructive interference. We further demonstrate that in small-angle twisted $MoTe_2/WSe_2$, the SHG polarization state is governed by the interplay between twist angle $\alpha$ and phase difference $\Delta\phi$, and can be mapped onto trajectories on the Poincaré sphere. At excitation energies satisfying $\Delta\phi + 3\alpha = 180°$, the SHG output becomes nearly circularly polarized (ellipticity ≈ 0.91) and undergoes an abrupt 90° azimuthal rotation, corresponding to a geometric polarization singularity in the parameter space. Our findings open new routes for exciton-resonance engineered nonlinear photonics and stacking-resolved optical functionality in moiré materials.

## I. INTRODUCTION

In TMDC heterobilayers, the moiré superlattices formed by small lattice and/or angle mismatch can host moiré trapped excitonic states, strong correlation effects and emergent topological phenomena[1-4]. The stacking configuration (AA or AB) modulates the moiré potential landscape and interlayer electronic hybridization, which together govern the band topology of the moiré system. Theoretical studies have shown that in nearly AA-stacked TMDC moiré bilayers, interlayer coupling with nontrivial real-space texture can give rise to topologically nontrivial Chern bands[5,6]. By contrast, in nearly AB-stacked TMDC moiré bilayers, the relevant valence bands of the two layers carry opposite spins near the same K or K' point in momentum space, leading to an interlayer decoupled topologically trivial state. Consistent with this picture, integer and fractional quantum anomalous Hall (QAH) effects have been observed exclusively in nearly AA-stacked TMDC homobilayers[7-10]. However, an apparent exception arises in $MoTe_2/WSe_2$ heterobilayers, in which the stacking configuration previously assigned as "AB" hosts a QAH state[11-24], whereas the structure labeled as "AA" was reported to be topologically trivial[11,17,25]. To reconcile this discrepancy, several theoretical scenarios have been proposed, including lattice reconstruction in the nominally "AB"-stacked $MoTe_2/WSe_2$ that has strain-induced pseudomagnetic fields[20] or breaks out-of-plane spin conservation[17], both of which can stabilize a valley-polarized QAH ground state[17-20]. Alternatively, other studies have proposed a valley-coherent QAH ground state driven by excitonic insulating gap arising from interlayer Coulomb interactions[16,21-24]. Therefore, the microscopic origin of the QAH state in this system remains highly debated.

The identification of AA or AB stacking $MoTe_2/WSe_2$ has commonly relied on their SHG response: 0° (also referred to as *H*- or AA-) stacking yields constructive SHG interference, whereas 60° (also referred to as *R*- or AB-) stacking results in destructive interference and a much weaker SHG intensity. This convention originates from the analogies with homobilayers, where *H*-stacked TMDC and hexagonal boron nitride (*h*-BN)[26-29] bilayers exhibit strong suppression of SHG, while *R*-stacked TMDC[30] and BN[31] bilayers show enhanced SHG responses. This method has subsequently been applied, often without further scrutiny, to identify AA or AB stacking in other heterobilayers, such as $WSe_2/WS_2$[32-34], $WSe_2/MoS_2$[35,36], and $MoS_2/WS_2$[37]. In these systems, however, excitonic resonances can strongly influence both the magnitude and phase of the SHG response of each individual layer[37,38], potentially complicating a direct analogy with homobilayer behavior.

Here, we show that the previous assignment of "AA" and "AB" -stacked $MoTe_2/WSe_2$ based on SHG measurements needs to be reconsidered. This misassignment arises from the overlooking of a nearly $\pi$ phase difference ($\Delta\phi \approx 180°$) in the second-order nonlinear susceptibility $\chi^{(2)}$ between the $WSe_2$ and $MoTe_2$ layers, which originates from their distinct excitonic resonances. These results demonstrate that only AA-stacked $MoTe_2/WSe_2$ hosts topologically nontrivial bands and the QAH state, restoring consistency with the original theoretical expectations[5,6] and obviating the need to invoke more exotic mechanisms[11-24]. Furthermore, we experimentally observe that tuning the excitation energy across a critical phase condition drives the SHG output to near-circular polarization accompanied by an abrupt 90° rotation of its polarization azimuth, signaling a geometric polarization singularity.

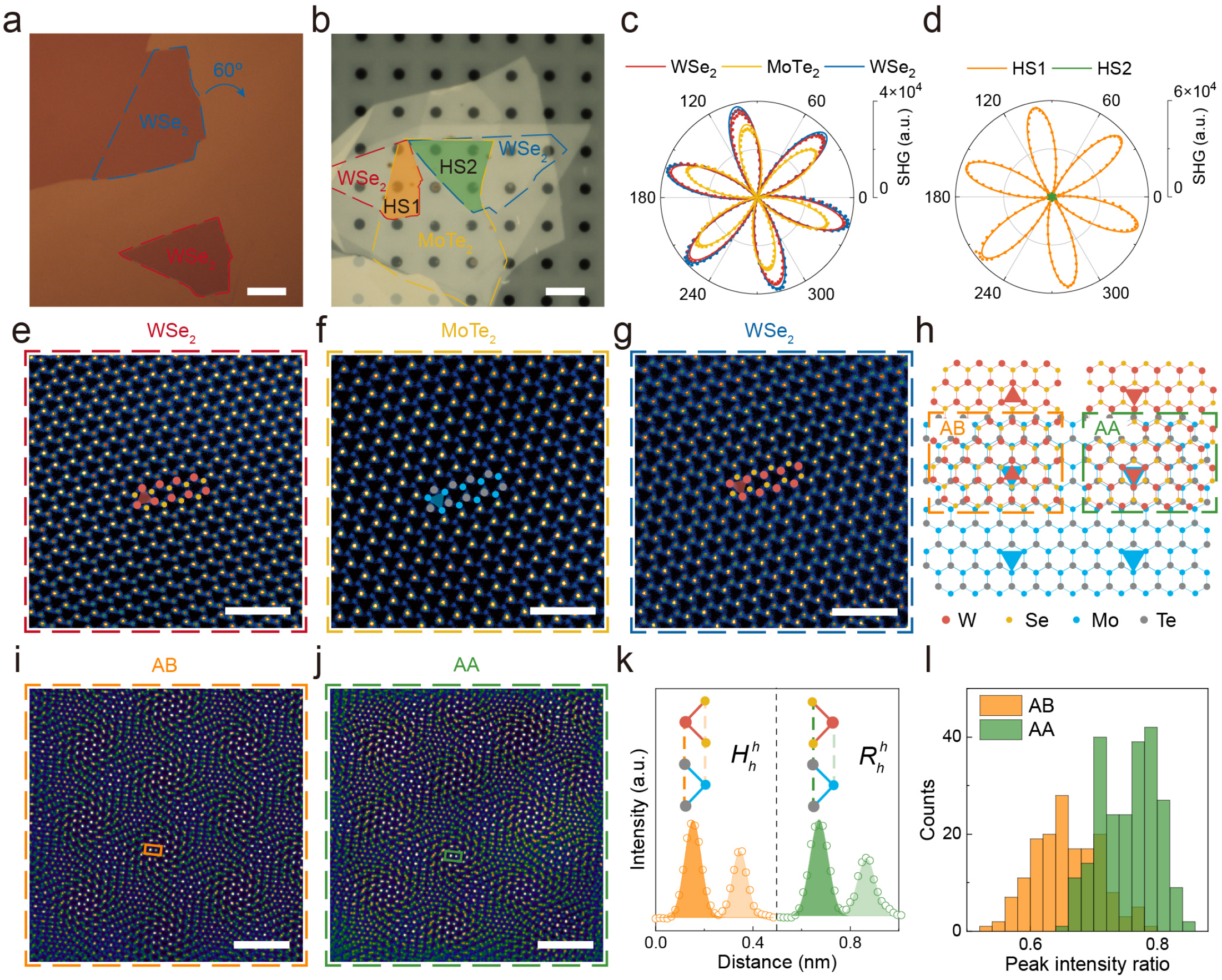

FIG. 1. Identifying AA and AB stacking of $MoTe_2/WSe_2$ by multislice electron ptychography (MEP). (a) Optical micrograph of two isolated monolayer $WSe_2$ flakes (outlined in red and blue) with identical crystal orientation, which are used to construct the heterostructure in (b). These flakes are sequentially picked up with a relative twist angle of 60° and stacked in a non-overlapping configuration. A subsequent $MoTe_2$ flake is aligned and overlapped with the two $WSe_2$ flakes using angle-resolved SHG before final assembly. (b), Optical micrograph of fully assembled device D1, encapsulated between top and bottom *h*-BN flakes and released onto a $SiN_x$ membrane with 2 μm holes for MEP measurements. Red, yellow, and blue dashed outlines mark the monolayer $WSe_2$, $MoTe_2$, and 60°-twisted $WSe_2$ flakes, respectively. Orange and green shaded areas indicate the two heterostructure regions (HS1 and HS2). Angle-resolved SHG of (c) three monolayer regions and (d) two heterostructure regions. Colored labels match those in (b). e-g, ADF image of I $WSe_2$, (f) $MoTe_2$, (g) 60°-twisted $WSe_2$. MEP image of (i) HS1, (j) HS2. H, Schematic illustration of the stacking configuration between $MoTe_2$ in (f) and $WSe_2$ in I and (g). Red and blue triangles represent the orientations of the W and Mo atomic arrangements within the hexagonal unit cells, respectively. k, Intensity profile across the orange and green frame in (i) and (j), corresponding to $R_\mathrm{h}^\mathrm{h}$ or $H_\mathrm{h}^\mathrm{h}$ region. L, Histograms of peak ratio measured in $R_\mathrm{h}^\mathrm{h}$ or $H_\mathrm{h}^\mathrm{h}$ region in (i) and (j). Scale bars, 10 μm (a-b), 1 nm (e-g), 2 nm (i-j).

In monolayer TMDCs, the angle-resolved SHG intensity $I_{\|}$ under the parallel configuration (see Appendix B) follows $I_{\|} = I_0 \cos^2(3\theta_\omega)$, where $\theta_\omega$ is the angle between the armchair (AC) direction and the polarization of the pump beam, and $I_0 = I_{max}$ denotes the maximum SHG intensity. This configuration yields a sixfold angular pattern, with the minimum intensity $I_{min}$ approaching zero. Consequently, by identifying the AC direction of each monolayer from the SHG intensity maxima and aligning them accordingly, one can fabricate angle aligned TMDC moiré heterobilayers. However, due to the lack of SHG phase information, this method typically yields an arbitrary AA or AB stacking configuration[39]. We develop a strategy to deterministically create both AA and AB stacked TMDC moiré heterobilayers within a single device. As illustrated in Fig. 1(a) and schematically shown in Fig. S5, our approach involves sequentially picking up two nearby $WSe_2$ monolayers (outlined in red and blue) with identical crystal orientation, applying a 60° rotation between pickups, and then overlaying the assembly on to an angle-aligned $MoTe_2$ monolayer. The entire stack is encapsulated between thin *h*-BN layers to avoid sample degradation. This process yields one heterostructure region (HS1) with either AA or AB stacking, while the second region (HS2) necessarily adopts the complementary configuration [Fig. 1(b)]. Importantly, this design ensures an identical dielectric environment for both stacking configurations, providing a well-controlled experimental platform for direct comparison. Figs. 1(c) and 1(d) display the angle-resolved SHG (under ~780 nm excitation) results from different regions in the final device D1, including three nearly aligned monolayer regions ($WSe_2$, $MoTe_2$, and 60°-rotated $WSe_2$) and two heterostructure regions (HS1 and HS2). Compared to the monolayers, the SHG intensity is strongly enhanced in HS1 and substantially suppressed in HS2, which according to previous studies, would be assigned to "AA"[25] and "AB"[11-16] stacking, respectively.

**II. ATOMIC CONFIGUTATIONS**

Scanning transmission electron microscopy (STEM) annular dark-field (ADF) images are routinely used to resolve the atomic structures of two-dimensional materials. However, this method has large limitations in resolving power including image contrast and interpretability for the *h*-BN-encapsulated structure (such as Device D1 here) due to the projection and multiple scattering effects of the beam electrons. To directly determine the stacking orders corresponding to the two SHG features, we employ multislice electron ptychography (MEP)[40,41]. This technique utilizes multiple position-dependent diffraction patterns to reconstruct three-dimensional structures of samples at an ultrahigh lateral resolution and sub-3 nm depth resolution, largely eliminating the structural mixture of different stacking layers. The contrast of MEP images (phase of the transmission function) is linearly dependent on the number of atoms along the projection, with atomic number ($Z$) se nsitivity proportional to ~ $Z^{\gamma}$ where $\gamma$ is usually less than 0.7 and dependent on many imaging parameters[40]. We first perform MEP imaging from the monolayer samples near the heterostructures. In Figs. 1(e) and 1(g), the higher-contrast columns correspond to W ($Z$ = 74, marked by red spheres), while the lower-contrast columns represent two Se ($Z$ = 34, marked by yellow spheres, 2Se). In Fig. 1(f), the higher-contrast columns correspond to two Te ($Z$ = 52, marked by gray spheres, 2Te), and the lower-contrast columns represent Mo ($Z$ = 42, marked by blue spheres). The contrast difference is consistently reproduced in the simulations (Supplemental Material, Fig. S1[42]). Angle-resolved SHG cannot distinguish between two $WSe_2$ monolayers differed by 60° [Fig. 1(c)]. This limitation is resolved by determining the orientation of the metal atoms within the hexagonal unit cell. As shown in Fig. 1(h), the tungsten atoms from the two $WSe_2$ layers form triangular lattices oriented at 60° to each other and are stacked with the $MoTe_2$ layer at angles close to 60° and 0° (determined by angle-resolved SHG to be 56° and 3.8°), respectively. Therefore, based on the directly resolved monolayer orientations, we can safely come to a counterintuitive conclusion that HS1 and HS2 correspond to nearly AB- and AA-stacked configurations, respectively.

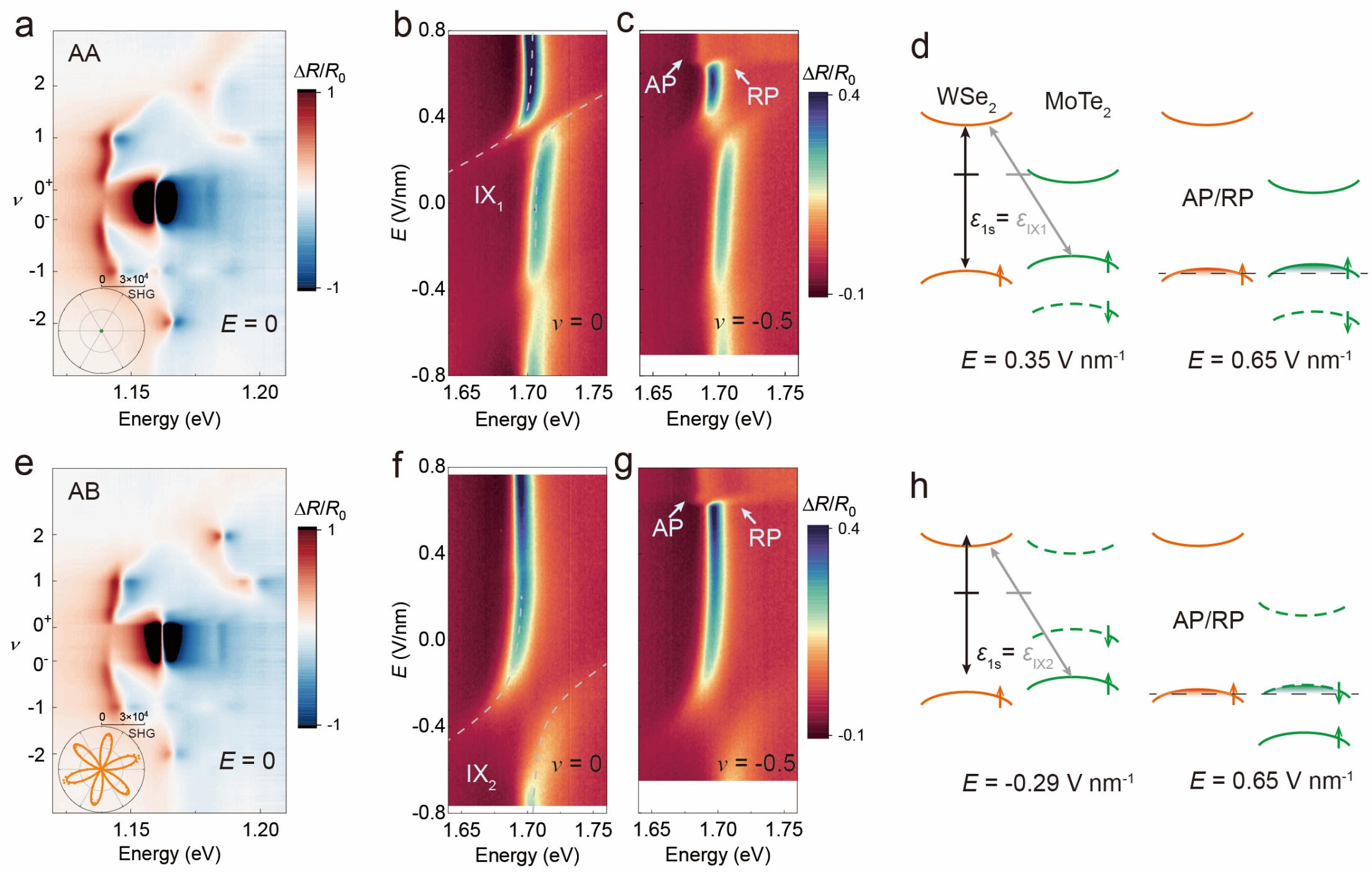

FIG. 2. Exciton hybridization and band-alignment in AA or AB stacking $MoTe_2/WSe_2$. (a), (e), Doping-dependent reflectance contrast spectra at zero electric field ($E = 0$) near the energy of $MoTe_2$ $A_{1s}$ exciton in (a) AA-stacked and (e) AB-stacked regions. Insets: angle-resolved SHG intensity used to identify the stacking configurations. Electric-field-dependent reflectance contrast near the energy of $WSe_2$ $A_{1s}$ exciton in AA-stacked region at filling factors (b) $\nu = 0$, (c) $\nu = -0.5$, and in AB-stacked region at (f) $\nu = 0$, (g) $\nu = -0.5$. Dashed curves in (b) and (f) are the best fit to the two-level model. Arrows in (c) and (g) marks the hole-doped AP and RP. D, h, Schematic of interlayer and intralayer exciton hybridization and band alignment for (d) AA and (h) AB stacking. Spin-up (spin-down) bands are denoted by solid (dotted) lines and upward (downward) arrows. $\varepsilon_{1s}$ denotes the energy of $WSe_2$ $A_{1s}$ exciton. $\varepsilon_{IX1}$ and $\varepsilon_{IX2}$ denotes the energy of $IX_1$ and $IX_2$. Application of a positive/negative electric field increases/decrease $\varepsilon_{IX1}$ and $\varepsilon_{IX2}$. Hybridization occurs when $\varepsilon_{1s} = \varepsilon_{IX1}$ or $\varepsilon_{1s} = \varepsilon_{IX2}$. A higher electric field Is required for band alignment due to the difference in exciton binding energies. The emergence of $WSe_2$ AP/RP features confirms charge transfer into $WSe_2$.

We further examine the MEP images of the two moiré heterobilayer regions. The atomic columns with the highest contrast in Fig. 1(i) and 1(j) are located within the $R_h^h$ or $H_h^h$ areas (Supplemental Material, Fig. S2[42]), where superscript and subscript $h$ represent the vertically aligned hexagon center for $WSe_2$ and $MoTe_2$. In the AB-stacked region [Fig. 1(k)], the vertically aligned atomic column pairs consist of W+2Te and 2Se+Mo. In contrast, the AA-stacked region exhibits combinations of 2Se+2Te and W+Mo. This indicates that the $H_h^h$ region in the AB stack exhibits higher atomic column contrast, as shown by the peak intensity ratio in Fig. 1(k). We further quantified the peak intensity ratio of analogous atomic columns within the $R_h^h$ or $H_h^h$ regions in Fig. 1(i) and 1(j). As summarized in Fig. 1(l), the AA region exhibits a significantly higher peak intensity ratio than the AB region, consistent with the simulated images (Supplemental Material, Fig. S3[42]). The contrast difference in the intermediate regions between Moiré centers is also consistent between experimental MEP images and corresponding simulations, which further verifies the AB and AA stackings (Supplemental Material, Fig. S3[42]). Therefore, our MEP results conclusively demonstrate that the $MoTe_2/WSe_2$ heterobilayers exhibit anomalous SHG interference—specifically, destructive interference in the AA-stacked region and constructive interference in the AB-stacked region, in stark contrast to conventional expectations.

## III. BAND ALIGNMENTS AND EXCITON HYBRIDIZATION

Beyond their atomic registries, AA- and AB-stacked TMDC heterobilayers also differ in their momentum-space band alignments, leading to contrasting exciton hybridization behaviors[25,43]. To investigate these effects, we fabricate a dual-gated device D2 incorporating both AA- and AB- stacked $MoTe_2/WSe_2$ heterobilayers with angle misalignment less than ~1° (see Appendix A), enabling independent control of the vertical electric field ($E$) and the carrier density. Hereafter, we label the region exhibiting destructive SHG [inset in Fig. 2(a)] as AA and the region exhibiting constructive SHG [inset in Fig. 2(e)] as AB. We first examine the doping-dependent reflectance spectra near the $MoTe_2$ $A_{1s}$ exciton (~1.16 eV) under a zero electric field ($E$ = 0 V nm$^{-1}$) [Figs. 2(a) and 2(e)]. Owing to the type-I band alignment, where $WSe_2$ has a higher conduction band minimum and a lower valence band maximum than $MoTe_2$, both hole and electron doping drive the neutral $MoTe_2$ $A_{1s}$ exciton into exciton-polaron states. This evolution is characterized spectroscopically by a redshift of the attractive polaron (AP) and a blueshift of the repulsive polaron (RP). In addition, pronounced modulations in both reflection contrast and resonance energy are observed at moiré band fillings of one ($\nu = \pm 1$) and two ($\nu = \pm 2$) electrons or holes per superlattice site. Interestingly, the doping-dependent spectral differences between the AA- and AB-stacked regions are less pronounced than those reported in $WSe_2/WS_2$ moiré systems[43,44], suggesting similar spatial distributions and localization of the excitonic wavefunctions in AA- and AB-stacked $MoTe_2/WSe_2$[45].

Figs. 2(b) and 2(f) present the electric-field-dependent reflectance spectra near the $WSe_2$ $A_{1s}$ exciton under zero doping ($\nu = 0$) for the AA- and AB-stacked regions, respectively. Notably, pronounced energy anticrossing behaviors between interlayer excitons (IXs, with slopes proportional to the out-of-plane dipole moment) and the $WSe_2$ $A_{1s}$ intralayer exciton (at 1.70 eV) are observed at opposite electric-field polarities in the two stackings. The linear increase of the IX energy with applied electric fields indicates that the IX originates from an electron in the conduction band of the top $WSe_2$ layer and a hole in the valence band of the bottom $MoTe_2$ layer. Crucially, the hybridization condition in each region follows the stacking-dependent selection rules expected for AA versus AB. In the AA-stacked region, the top valence bands of both layers become nearly aligned under a positive electric field ($E$ = 0.35 V $nm^{-1}$), enable spin-conserving hole tunneling and leading to hybridization between the $WSe_2$ $A_{1s}$ exciton and $IX_1$ [Figs. 2(b)]. In contrast, in the AB-stacked region, hole tunneling between the top valence bands of the two layers is spin-forbidden. Instead, clear exciton hybridization is only observed when the bottom valence band of $MoTe_2$ is nearly aligned with the top valence band of $WSe_2$ under a negative electric filed ($E$ = –0.29 V $nm^{-1}$) [Figs. 2(f)] Accordingly, $IX_1$ ($IX_2$) stems from transitions between the top (bottom) valence band of $MoTe_2$ and the conduction band of $WSe_2$, as illustrated schematically in the left panels of Fig. 2(d) [Figs. 2(h)].

We model the coupled excitonic system using a two-level Hamiltonian (see Appendix E), with the gray dashed curves in Figs. 2(b) and 2(f) representing the best fits to the experimental data. The coupling strengths $|W_i|$, corresponding to the hybridization between $IX_i$ (i = 1, 2) and the intralayer exciton, are extracted to be $|W_1|$ = 13 meV and $|W_2|$ = 25 meV. Both values exceed the intralayer exciton linewidth (~10 meV), indicating that the exciton hybridization of the two stackings are both in a strong coupling regime. The larger value of $|W_2|$ ($>|W_1|$) is likely due to a closer interlayer spacing in AB stacking[46]. In the AA-stacked region [Figs. 2(b)], an additional weak anticrossing is observed at $E$ = –0.38 V $nm^{-1}$, which may arise from small finite hybridization between a spin-triplet IX and the intralayer exciton. Furthermore, the energy of $IX_1$ at $E$ = 0 V $nm^{-1}$ is ~200 meV lower than that of $IX_2$, consistent with the energy scale of valence band spin-orbit splitting in $MoTe_2$. This observation further supports the band assignments of $IX_1$ and $IX_2$ shown in Figs. 2(d) and 2(h). Taken together, the distinct exciton hybridization behaviors observed in Figs. 2(b) and 2(f) provide an independent confirmation of the stacking configuration in $MoTe_2/WSe_2$ heterobilayers.

Under slight hole doping ($\nu$ = –0.5), the charge carriers are fully distributed in the $MoTe_2$ layers for $E$ < ~0.65 V $nm^{-1}$, as indicated by the strong resonance of the $WSe_2$ $A_{1s}$ exciton in Fig. 2c and 2g. Increasing the electric field beyond ~0.65 V $nm^{-1}$ reveals a hallmark of charge transfer from $MoTe_2$ to $WSe_2$: the emergence of AP and RP associated with the $WSe_2$ A 1s exciton, indicated by white arrows in Fig. 2c and 2g [band alignment schematics shown in the right panels of Figs. 2(d) and 2(h)]. A previous study[25] attributed a hybridization feature observed at $E$ = 0.32 V $nm^{-1}$ to band alignment in “AB”-stacked $MoTe_2/WSe_2$. Our comparative measurements, however, demonstrate that this feature instead arises from energy degeneracy between interlayer excitons and the intralayer $A_{1s}$ exciton. Because the binding energy of interlayer excitons is several tens of meV smaller than that of intralayer excitons[47-49], a substantially larger electric field is required to achieve true band alignment, as directly evidenced in Figs. 2(c) and 2(g).

## IV. EXCITATION ENERGY DEPENDENT SHG INTENSITY AND PHASE

In the first two sections, we establish a direct correspondence between the stacking configuration (AA or AB) and SHG interference (destructive or constructive) in $MoTe_2/WSe_2$ at a fixed excitation wavelength of 780 nm. Intriguingly, this interference behavior persists over a broad excitation-energy range. To investigate this behavior in detail, we perform excitation-energy-dependent SHG measurements at fixed excitation power of ~1 mW for AA and AB stacking configurations, as well as individual monolayer $WSe_2$ and $MoTe_2$ within the same device. Fig. 3(a) displays the results for monolayer $WSe_2$ and $MoTe_2$ over an excitation energy range of 1.15-1.70 eV. In $WSe_2$, a broad SHG resonance is observed around 1.47 eV, in good agreement with previous experimental reports[50,51], whereas in $MoTe_2$, the most prominent resonance peak appears near 1.25 eV. In parallel, we carry out first-principles calculations based on GW plus Bethe-Salpeter equation (GW-BSE) framework to provide a theoretical comparison and gain further insight. The overall line shapes of these SHG excitation spectra are qualitatively reproduced by our calculated SHG susceptibility square, $|\chi^{(2)}|^2$, obtained using the exciton-state coupling (ESC) formalism[52] (see Appendix F). Through detailed analysis, we identify that the dominant SHG resonance peaks in the two materials arise from two-photon resonances with the C exciton in $WSe_2$ and the D exciton in $MoTe_2$, which are accordingly assigned as $C_{1/2}$ and $D_{1/2}$, respectively. The C exciton in $WSe_2$ originates from band-nesting regions of the Brillouin zone, whereas the D exciton of $MoTe_2$ consists of electron-hole pairs around the M point, as illustrated in the inset of Figs. 3(b) and 3(c). Near the energies of the C and D excitons, two-photon resonances with additional exciton states also contribute to SHG, resulting in the broad spectral features observed in Fig. 3(a).

We now extend our analysis to the SHG spectra of AA- and AB-stacked $MoTe_2/WSe_2$. As shown in Fig. 3(d), the measured SHG intensity from the AB region is consistently stronger than that from the AA region across the entire excitation-energy range of 1.15-1.70 eV. For comparison, Fig. 3(e) presents the calculated $|\chi^{(2)}|^2$ spectra for AA- and AB-stacked $MoTe_2/WSe_2$. These spectra are obtained by squaring the sum (for AA stacking) or difference (for AB stacking) of the complex-valued $\chi^{(2)}$ from the individual monolayer $WSe_2$ and $MoTe_2$ ((see Appendix F), under the assumption that interlayer coupling effects are weak. We find both the energy-dependent trends and the intensity contrast (AB > AA) generally agree with the experimental results across most of the spectral range, supporting the validity of this approximation. The deviations below ~1.2 eV might stem from the discrepancies between the calculated and experimental $\chi^{(2)}$ of monolayer $WSe_2$ in this low-energy region [upper panel of Fig. 3(a)].

We propose that the reduced SHG intensity of the AA stacking relative to the AB stacking can be understood from the phase difference $\Delta\phi = \phi_W - \phi_{Mo}$ between $WSe_2$ and $MoTe_2$, where $\phi_W$ and $\phi_{Mo}$ denote the phases of their SHG susceptibilities $\chi^{(2)}$. When $\Delta\phi$ lies between 90° and 270°, the two layers interfere destructively in the AA stacking but constructively in the AB stacking. This is indeed the case, as reflected in the lower panel of Fig. 3(f), where the cosine of the phase difference is plotted. Moreover, as shown in the upper panel of Fig. 3(f), the AA/AB SHG intensity ratio remains below 5% between 1.51 eV and 1.61 eV, indicating a strong suppression of SHG in the AA-stacked region. Across this spectral range, $\cos(\Delta\phi)$ approaches −1 [lower panel, Fig.

3(f)], corresponding to a phase difference close to 180°. This unconventional interference behavior arises from the involvement of the different types of exciton states, namely D excitons in $MoTe_2$ and C excitons in $WSe_2$, as discussed above. By contrast, in conventional heterostructures such as $WSe_2/WS_2$, the SHG responses from the two layers are governed by the same exciton type (C excitons), giving rise to interference behavior distinct from that observed here[33,50,51].

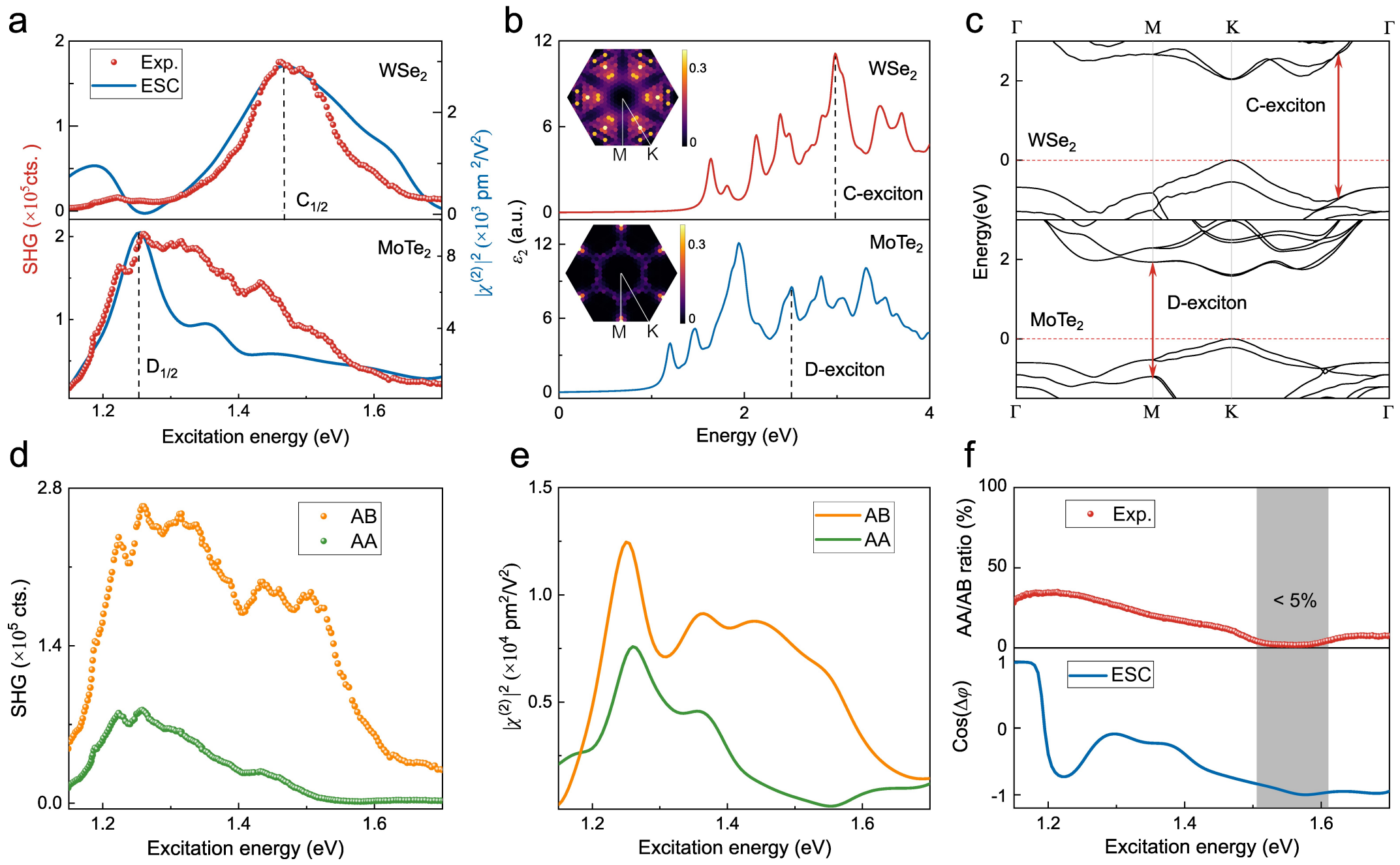

FIG. 3. Excitation-energy-dependent SHG intensity and phase difference in $MoTe_2/WSe_2$. (a) Upper panel: Experimentally measured SHG intensity of $WSe_2$ (red dots) and the calculated $|\chi^{(2)}|^2$ from exciton-state-coupling (ESC) calculations (blue line) as a function of excitation photon energy. Lower panel: Corresponding data for $MoTe_2$. (b) Upper panel: Calculated imaginary part of the dielectric function ($\varepsilon_2$) for $WSe_2$ from GW-BSE. Inset: k-space envelope function of the C exciton. Lower panel: Calculated $\varepsilon_2$ for $MoTe_2$. Inset: k-space envelope function of the D exciton. (c) Upper and lower panels: GW quasiparticle band structures of $WSe_2$ and $MoTe_2$, respectively. Transitions contributing to the C and D excitons are indicated by red arrows. (d) Measured SHG intensity of AB- and AA-stacked $MoTe_2/WSe_2$ as a function of excitation photon energy. (e) Calculated $|\chi^{(2)}|^2$ for AB- and AA-stacked configurations as a function of excitation photon energy. (f) Upper panel: AA/AB SHG intensity ratio versus excitation photon energy. The spectral range for AA/AB intensity ratio below 5% is highlighted by the shaded region, which covers the wavelength commonly used in experiments (~800 nm). Lower panel: Cosine of the phase difference, cos $(\Delta\phi)$, derived from the calculated complex $\chi^{(2)}$ of the two layers. The values near $-1$ within the shaded area indicate that $\Delta\phi \approx 180°$.

## V. SHG POLARIZATION STATE MANIPULATION

We next analyze how the SHG polarization state in AA-stacked $MoTe_2/WSe_2$ evolves with three parameters: the polarization angle of the linearly polarized excitation laser $\theta_\omega$ (defined relative to the x-axis chosen along the bisector of the two armchair directions), the interlayer SHG phase difference $\Delta\phi$, and the relative twist angle $\alpha$ [schematic in Fig. 4(a)]. The total SHG field arises from the coherent superposition of the linearly polarized SHG fields emitted by the two layers $\vec{E}(2\omega) = \overrightarrow{E_1}(2\omega) + \overrightarrow{E_2}(2\omega)$, which generally produces an elliptically polarized output (see Appendix G and Supplemental Material Fig. S6[42]). To quantitatively describe the polarization state, we employ the normalized Stokes parameters $s_i = S_i/S_0$ ($i$ = 1, 2, 3), where $S_0$ denotes the total SHG intensity. The polarization state corresponds to a point on the Poincaré sphere, where $(s_1, s_2, s_3)$ define its Cartesian coordinates satisfying $s_1^2 + s_2^2 + s_3^2 = 1$. The polarization azimuth $\theta_{2\omega}$ (orientation of the semi-major axis relative to the x-axis) is determined by $\tan(2\theta_{2\omega}) = s_2/s_1$, while $s_3$ specifies the handedness and ellipticity (defined as the ratio between the semi-minor and semi-major axes of the polarization ellipse). Within our interference model, $s_3$ is calculated to be independent of the excitation polarization angle $\theta_\omega$, reflecting that the degree of ellipticity is governed solely by $\Delta\phi$ and $\alpha$.

Experimentally, these quantities can be extracted from the polarization-resolved SHG measurements performed in the parallel detection configuration, where the SHG field is projected onto the pump polarization direction. The measured intensity [schematic shown in Fig. 4(b)] follows:

$$I_\parallel \propto \left| Ae^{i2\omega t}\cos 3\left(\frac{\alpha}{2} - \theta_\omega\right) + Be^{i(2\omega t - \Delta\phi)}\cos 3\left(\frac{\alpha}{2} + \theta_\omega\right) \right|^2 \quad (1)$$

where $A$ and $B$ denote the SHG electric field amplitudes of $WSe_2$ and $MoTe_2$, respectively. As $\theta_\omega$ is rotated, the SHG polarization ellipse is projected along different directions, yielding a characteristic six-fold angular modulation pattern. For device D2 ($\alpha$ = 1°), polarization-resolved SHG measurements are performed over an excitation-energy range of 1.512 to 1.560 eV [Fig. 4(e)]. The maximum and minimum intensities $I_{max}$ and $I_{min}$ occur when the detection axis aligns with the major and minor axes of the ellipse, respectively. The polarization ellipticity is therefore extracted as $\sqrt{I_{min}/I_{max}}$ [second panel of Fig. 4(e)]. Notably, upon tuning the excitation energy across 1.533 eV, we observe an abrupt 30° phase shift of the six-fold angular pattern [see representative measurements at 1.523 eV and 1.550 eV in Figs. 4(c) and 4(d), respectively]. Simultaneously, the ellipticity reaches a maximum value of approximately 0.91, signifying a near-circular polarized state.

To elucidate this behavior, we consider $A = B$, which is justified by the comparable SHG amplitudes of the two layers within 1.321 to 1.565 eV (Supplemental Material, Fig. S10[42]). Eq. (1) then simplifies to:

$$I_\parallel \propto A^2[1 + \cos\Delta\phi\cos 3\alpha + \cos 6\theta_\omega(\cos\Delta\phi + \cos 3\alpha)] \quad (2)$$

The angular modulation term vanishes when $\cos\Delta\phi + \cos 3\alpha = 0$, or equivalently $\Delta\phi + 3\alpha = 180°$. Under this condition, the SHG response becomes independent of $\theta_\omega$, corresponding to circular polarization. For $\Delta\phi < 180° - 3\alpha$, the intensity maxima occur at $\theta_\omega = 0°$ (mod 60°) with the polarization azimuth satisfying $\theta_{2\omega} = -2\theta_\omega$ (mod 180°). For $\Delta\phi > 180° - 3\alpha$, the maxima shift to $\theta_\omega = 30°$ (mod 60°) and the azimuth becomes $\theta_{2\omega} = -2\theta_\omega + 90°$ (mod 180°). This interchange produces the observed 30° shift in the six-fold angular pattern and corresponds to a 90° rotation of the polarization azimuth $\theta_{2\omega}$. As shown in Fig. 4(f), the full evolution of $I_\parallel$, ellipticity, $\theta_{2\omega}$, and polarization state (Pol.) in the parameter space ($\theta_\omega$, $\Delta\phi$, $\alpha$) is quantitatively reproduced by simulations (Also see varying $\alpha$ in Supplemental Material Fig. S7[42]).

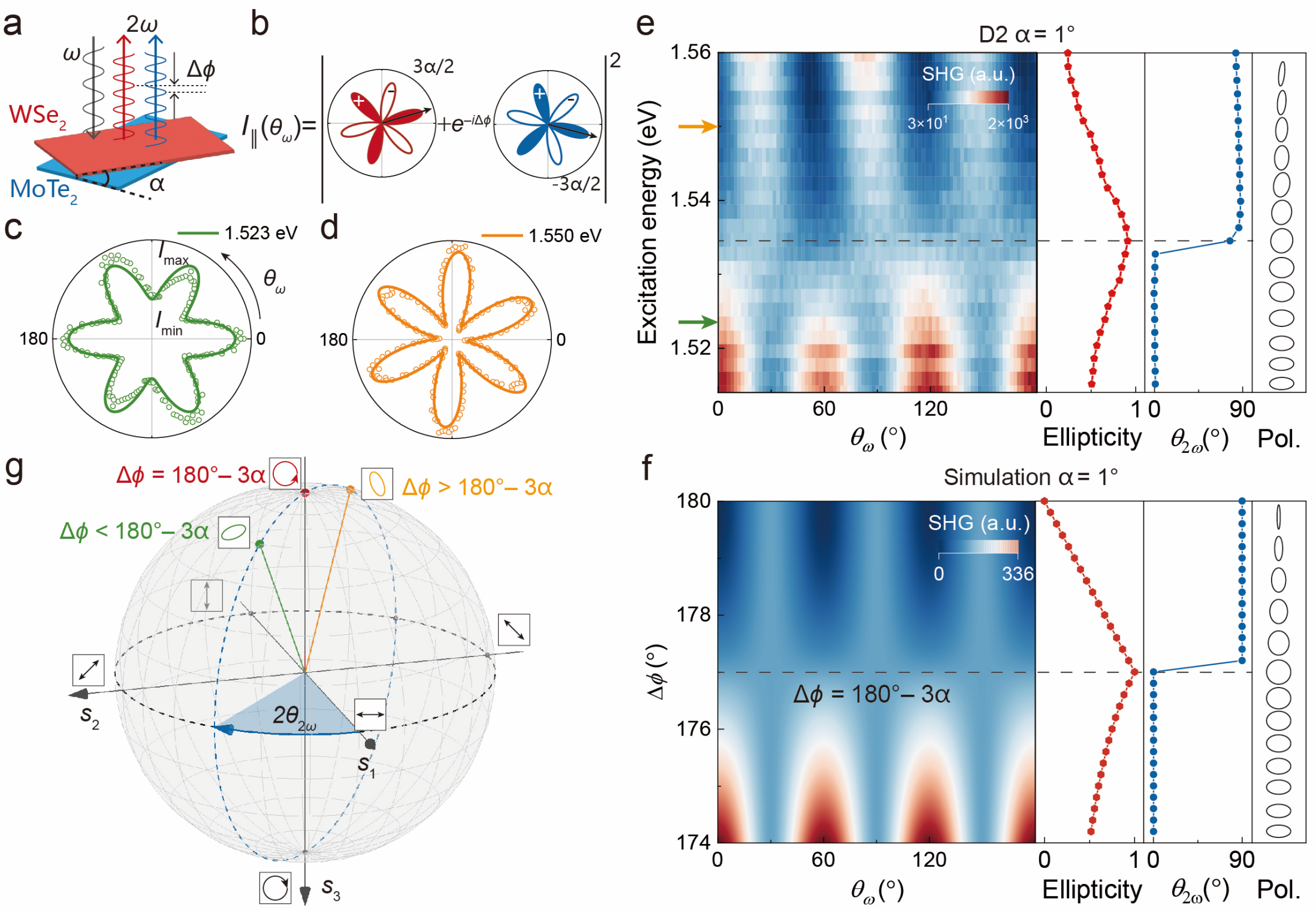

FIG. 4. SHG polarization state manipulation in nearly AA stacked $MoTe_2/WSe_2$. (a) Schematics of SHG superposition from $WSe_2$ and $MoTe_2$ with a phase difference $\Delta\phi$ and twist angle $\alpha$. (b) The coherent synthesis of $I_\parallel(\theta_\omega)$ from two SHG fields of the $WSe_2$ layer (red) and the $MoTe_2$ layer (blue), characterized by angle misalignment $3\alpha$ and phase difference $\Delta\phi$. The filled and empty lobes are contrasted to have opposite SHG fields. Representative polarization dependent SHG measurement of D2 ($\alpha = 1°$ region) with the excitation energy of (c) 1.523 eV and (d) 1.550 eV, where SHG power maxima $I_{max}$ and minima $I_{min}$ can be extracted. (e) Left panel: SHG ($I_\parallel$) as a function of excitation energy and polarization, measured from D2 ($\alpha = 1°$ region). Right panels: SHG ellipticity, azimuthal angle $\theta_{2\omega}$ and polarization states (Pol.) as functions of excitation photon energy. (f) Left panel: simulated SHG as a function of $\Delta\phi$ and $\theta_\omega$, with $\alpha$ fixed at 1°. Right panel: SHG ellipticity, azimuthal angle $\theta_{2\omega}$ and Pol. as functions of $\Delta\phi$. (g) Schematic Poincaré sphere showing the change in polarization state near the south pole ($\Delta\phi = 180° - 3\alpha$). When the changes in $\Delta\phi$ and $\alpha$ drive the point

(from green to orange) on the sphere through the pole (marked in red), $\theta_{2\omega}$ changes by 90°. The normalized Stokes parameters $(s_1, s_2, s_3)$ define the Cartesian coordinates.

As shown in Fig. 4(g), the condition $\Delta\phi + 3\alpha = 180°$ corresponds to the polarization state reaching the south pole of the Poincaré sphere (left-handed circular polarization). At this pole, the polarization azimuth becomes undefined, resulting in a polarization singularity in parameter space. As $\Delta\phi$ (or $3\alpha$) varies across this condition [see the trajectory through the green–red–orange points in Fig. 4(g)], the polarization state crosses the pole along a meridian of the Poincaré sphere, leading to a discontinuous 90° jump in the ellipse azimuth $\theta_{2\omega}$. Consistent with this picture, the experimentally observed ellipticity approaches ~0.91 and $\theta_{2\omega}$ shifts by ~87° near 1.533 eV, indicating proximity to the singular point satisfying $\Delta\phi \approx 180° - 3\alpha$. At excitation energies well below or above this resonance, the SHG output approaches linearly polarization, with orthogonal orientations on either side of the transition [see right panels of Figs. 4(e) and 4(f)]. Similar behavior is observed in another device D1 with $\alpha = 3.8°$ (Supplemental Material, Fig. S8[42]), whereas near-AB-stacked regions do not exhibit such polarization evolution. These results collectively demonstrate that the coherent superposition model incorporating interlayer phase difference $\Delta\phi$ effectively quantitatively captures the SHG polarization behavior in $MoTe_2/WSe_2$ heterobilayers.

## VI. CONCLUSIONS

To summarize, by resolving the stacking ambiguity in $MoTe_2/WSe_2$ and restoring the correspondence between optical stacking identification and topological moiré bands, our work establishes a unified framework linking excitonic resonances, nonlinear interference, and emergent band topology. We further demonstrate that a broad manifold of SHG polarization states can be continuously accessed by tuning the interlayer phase difference $\Delta\phi$ (via excitation energy) and the twist angle $\alpha$, with polarization trajectories that map directly onto the Poincaré sphere. This interlayer phase engineering enables deterministic control of nonlinear polarization in an atomically thin platform, offering a compact route towards tunable classical frequency-doubling sources and potentially phase-tailored quantum light generation.

## ACKNOWLEDGMENTS

Y.X. and Y.D.W. designed the scientific objectives and co-wrote the manuscript. Y.D.W. fabricated the devices and performed the optical reflectance measurements and analysis. Y.L. and Y.D.W performed the SHG measurements and analysis under the guidance of K.Q.L. and Y.X. C.S.C. and Y.Z.J. performed the MEP measurements and analysis under the guidance of Z.C. X.T.L performed the DFT calculations and GW-BSE under the guidance of J.W.R. S.L. grew the bulk $WSe_2$ crystals. All authors discussed the results and commented on the manuscript.

## APPENDIX A: DEVICE FABRICATION

The device is fabricated using a layer-by-layer dry pick-up method[53]. The $WSe_2$, $MoTe_2$, *h*-BN, and few-layer graphite are first mechanically exfoliated from bulk crystals onto $SiO_2$/Si substrates (285 nm oxidation layer). The thicknesses of these flakes are identified by their optical contrast. A thin film of polycarbonate on polydimethylsiloxane is employed as a stamp to pick up the flakes following the sequence shown in Supplemental Material Fig. S5[42]. Angle alignment of the two TMDC monolayers is assisted by angle-resolved SHG measurements (details described below). The complete stacks are then released at 180 °C onto $SiO_2$/Si or $SiN_x$/Si substrates.

The data presented in this paper are mainly collected from 2 devices. Supplemental Material Fig. S5[42] shows the device structures of D1 and D2, along with the angle-resolved SHG characterizations for D2 [results for D1 are shown in Figs. 1(c) and 1(d)]. D1 consists of $MoTe_2$/$WSe_2$ heterostructures with twist angles of 3.8° and 56°, encapsulated with thin *h*-BN (to enable higher atomic resolution) and transferred onto a $SiN_x$/Si substrate with a pre-etched hole for MEP measurements. D2 contains $MoTe_2$/$WSe_2$ heterobilayers with twist angles of 1° and 59.5°. It is a dual-gated device designed for optical reflection measurements, using thin *h*-BN as the dielectric layer (which withstands higher breakdown electric fields) and few-layer graphite as the top gate, bottom gate, and contact electrodes. The entire structure is released onto a $SiO_2$/Si substrate (with a 285 nm oxide layer) with a pre-patterned Au/Ti electrode. The processes involving exposed $MoTe_2$ flakes are carried out in a glove box.

**APPENDIX B: SHG MEASUREMENTS**

Angle-resolved SHG is employed to determine the crystal orientations of the TMDC layers. The excitation source is a fiber laser (Menlo Systems, ELMO 780) with a center wavelength of 780 nm, a repetition rate of 100 MHz, and a pulse duration of 100 fs. An excitation beam with 1.5 mW average power is focused normally onto the sample through a microscope objective with a numerical aperture (NA) of 0.6, resulting in a spot diameter of approximately 2 μm. The SHG signal is collected in reflection geometry using the same objective and detected by a spectrometer coupled to a CCD camera. A short pass filter is used in the detection path to block the fundamental wavelength. The excitation light first passed through a polarizer, followed by a half-wave plate used to control the polarization direction relative to the sample orientation. The reflected SHG signal passes through the same half-wave plate and an analyzer, which is kept parallel to the excitation polarization (i.e., parallel configuration). The SHG from monolayer TMDC exhibits a six-fold symmetry pattern, with maximum intensity occurring when the polarization aligns with the armchair direction of the crystal. The twist angles in the heterostructures of D1 and D2 are determined by comparing the SHG intensities from the exposed monolayer regions and the stacked regions.

To measure the excitation-energy-dependent SHG, a tunable Ti:Sapphire laser (Coherent, Chameleon Ultra II) with a pulse duration of 140 fs and a repetition rate of 80 MHz is used as the excitation source. The incident power is set with a motorized half-wave plate and a fixed Glan–laser polarizer, and monitored by a power meter. The beam is focused onto the sample with a 0.6-NA objective (Olympus LUCPLFLN 40×), giving a ~2 μm spot. The reflected SHG is collected by the same objective, and the fundamental is suppressed with a 680 nm short-pass filter (Semrock FF01-680/SP-25). The SHG is dispersed by a spectrometer (Princeton Instruments HRS 300, 150 grooves/mm grating) and recorded with a CCD camera (Princeton Instruments PIXIS 100). The excitation wavelength is scanned from 730 to 1080 nm in 5 nm steps while keeping the power constant at 1 mW.

For the energy- and polarization-dependent SHG measurements (Fig. 4), we add a superachromatic half-wave plate (B. Halle RSU 1.2.10) to rotate the relative angle between the crystal and laser polarization, and a linear polarizer (Thorlabs WP25M-UB) to analyze the SHG polarization. The half-wave plate is placed between the beamsplitter and the objective; the SHG passes back through the same plate, is filtered by the 680 nm short-pass edge filter, and is detected in the horizontal polarization channel set by the analyzer polarizer. The excitation wavelength is scanned from 720 to 930 nm in 10 nm steps at 1 mW. At each wavelength, the half-wave plate is rotated from 0° to 90° in 0.5° increments to vary the excitation polarization relative to the fixed crystal axes. Repeated scans confirm signal stability.

## APPENDIX C: MEP MEASUREMENTS AND SIMULATIONS

The four-dimensional STEM (4D-STEM) datasets are acquired using a probe aberration-corrected JEOL NEOARM200 electron microscope with 76-pA beam current, 200-keV beam energy, and 28.2-mrad aperture size, with the dose limited by the radiation resistance of the sample. A fast detector (Detrics Arina) with 192 × 192 pixels is used with an exposure time of 34 µs per frame and 400 × 400 diffraction patterns (scan step size of 0.50 Å). The experiments employ a defocused electron probe positioned approximately 20 nm above the sample's top surface. The original 192×192-pixel diffraction patterns are binned down to 96 ×96 pixels before ptychographic reconstruction. Reconstructions are performed with a slice thickness of 0.5 nm using multislice electron ptychography algorithms based on the maximum likelihood[54] and mixed-state probe[55,56]. To enhance real-space sampling, the diffraction patterns are padded to 144 pixels. Fourier frequencies from *h*-BN lattices are filtered out in Figs. 1(i) and 1(j) to eliminate the structural mixture of encapsulated *h*-BN from $MoTe_2/WSe_2$ heterostructures (see Supplemental Material Fig. S4[42] for more details). Simulated ADF and MEP images for single-layer $WSe_2$, $MoTe_2$, and $MoTe_2/WSe_2$ AA and AB stacking are calculated by using abTEM software[57]. Experimental parameters, including beam energy and probe-forming semi-angle, are applied in the simulations. For MEP simulations, the probe is focused 15 nm above the sample, and 97 ×81 diffraction patterns are generated with a step size of 0.5 Å. Each pattern has 192 ×192 pixels at 0.55 mrad per pixel same as the experimental parameters. Similar reconstruction parameters are also used to reconstruct the MEP phase images. Simulated ADF images are calculated under a collection angle 50-200 mrad and an illumination dose of $5 \times 10^5$ e/Å$^2$, similar to the experimental imaging condition.

## APPENDIX D: ELECTRIC-FIELD- AND DOPING-DEPENDENT REFLECTION CONTRAST SPECTRUM

The device D2 is mounted in a closed-cycle cryostat (Attocube, Attodry 2100) for reflection spectroscopy measurements at temperatures as low as 1.7 K. A halogen lamp served as the white light source for reflection measurements. The output from the lamp is collected by a single-mode optical fiber, collimated by a 10× objective, and then focused onto the sample using an objective with a numerical aperture (NA) of 0.8. The beam diameter on the sample is approximately 1 μm, with a power maintained below 1 nW. The reflected light is collected by the same objective and detected by a spectrometer coupled to a CCD camera.

The reflection contrast spectrum is obtained by comparing the reflected light intensity from the sample I with that from a featureless substrate region ($R_0$), defined as $(R - R_0)/R_0$, with a measurement sensitivity of approximately 0.1%. For reflection contrast spectra in the $MoTe_2$ $A_{1s}$ exciton region, a superluminescent diode (Exalos, EXS210007-01) with a center wavelength of 1070 nm and a full width at half maximum (FWHM) of 90 nm is used as the light source.

For device D2, the out-of-plane electric field is calculated using $E = (V_{bg}/t_{bot} - V_{tg}/t_{top})/2$, where $t_{top}$ = 9 nm and $t_{bot}$ = 13 nm represent the thicknesses of the top and bottom *h*-BN dielectric layers, respectively.

## APPENDIX E: MODELLING THE EXCITON HYBRIDIZATION

According to our electric field-dependent reflection spectrum, there are exciton hybridizations between the two interlayer excitons and the intralayer exciton, which occurs under positive and negative electric fields, respectively. To model the coupled excitonic system of AA and AB region, we employed a two-level Hamiltonian:

$$H = \begin{pmatrix} \varepsilon_{IXi} & W_i \\ W_i^* & \varepsilon_{1si} \end{pmatrix}, \tag{3}$$

the energy of an interlayer exciton depends linearly on the electric field $E$, given by $\varepsilon_{IXi} = \varepsilon_0 + D_i * E$, where $\varepsilon_0$ is its energy at zero field and $D_i$ is the out-of-plane dipole moment ($i$=1 or 2). Here, $\varepsilon_{IX1}$ and $\varepsilon_{IX2}$ denote the two interlayer exciton states, while $W_1$ and $W_2$ represent their respective coupling strengths to the $WSe_2$ intralayer $A_{1s}$ exciton.

In AA stacking region, the best fitting parameters are: $\varepsilon_{IX1} = 1.598$ eV, $\varepsilon_{1s} = 1.705$ eV, $D_1 = 0.31$ $e$·nm, $|W_1|$ = 13 meV. In AB stacking region, the best fitting parameters are: $\varepsilon_{IX2} = 1.780$ eV, $\varepsilon_{1s} = 1.700$ eV, $D_2 = 0.28$ $e$·nm, $|W_2|$ = 25 meV.

## APPENDIX F: FIRST PRINCIPLES CALCULATIONS

### 1. Excitons and $\chi^{(2)}$ within GW-BSE framework

Our first-principles calculations of the electronic structure of monolayer $WSe_2$ and monolayer $MoTe_2$, which serve as the starting point for the subsequent GW-BSE and ESC studies, are conducted using density functional theory (DFT) as implemented in the Quantum ESPRESSO package[58]. The Perdew-Burke-Ernzerhof (PBE) exchange-correlation functional and norm-conserving pseudopotentials[59] are employed in our calculations. The Kohn-Sham orbitals are constructed with a plane-wave basis with an energy cut-off of 80 Ry, with SOC fully incorporated. To avoid spurious interactions between periodic images in the supercell calculation, we included a 17 Å and a 15 Å vacuum thickness for $WSe_2$ and $MoTe_2$, respectively. The lattice constants after the relation are 3.32 Å for $WSe_2$ and 3.55 Å for $MoTe_2$.

To accurately capture the excitonic effect, we performed GW-BSE calculations on top of DFT using the BerkeleyGW package[60-62]. More specifically, by solving the following BSE equation,

$$[\varepsilon_c(\mathbf{k}) - \varepsilon_v(\mathbf{k})]A^S_{vc\mathbf{k}} + \sum_{v'c'\mathbf{k}'} K_{vc\mathbf{k}v'c'\mathbf{k}'} A^S_{v'c'\mathbf{k}'} = \Omega^S A^S_{vc\mathbf{k}} \tag{4}$$

we obtained the excitonic state $|S\rangle$ with energy $\Omega^S$, and its **k**-space envelope function $A^{(S)}_{vc\mathbf{k}}$. Here, $\varepsilon_n(\mathbf{k})$ is GW quasiparticle energy and $K_{vc\mathbf{k}v'c'\mathbf{k}'}$ is electron-hole interaction kernel matrix elements. Since obtaining well-converged GW quasiparticle energies is computationally demanding, we adopted values from previous studies[63,64] and interpolated them onto a 24×24×1 **k**-grid for the BSE calculations. The interaction kernel of BSE is constructed using a same **k**-grid, together with 12 valence band states 8 conduction band states. After getting the exciton states and energies, the imaginary part of dielectric function $\epsilon_2(\omega)$ incorporating excitonic effect are computed using $\epsilon_2(\omega) = 16\pi^2 e^2 \sum_S |\mathbf{e} \cdot \mathbf{R}_{S0}|^2 \delta(\hbar\omega - \Omega^S)$, where $R_{S0}$ are the optical transition matrix elements between an exciton state $|S\rangle$ and the ground state $|0\rangle$. In the calculation, a broadening parameter of 50 meV is used for results shown in Fig. 3.

The SHG susceptibility tensor of monolayer $MoTe_2$ and monolayer $WSe_2$ are computed based on the ESC formalism[52], which is given as

$$\begin{aligned}
\chi^{\mu,\nu\lambda}(2\omega;\omega,\omega) &= \frac{e^3}{2\epsilon_0 V}\sum_{n,m} \frac{R^{\mu}_{0n}R^{\nu}_{nm}R^{\lambda}_{m0}}{(2\hbar\omega - \Omega_n + i\eta_n)(\hbar\omega - \Omega_m + i\eta_m)} \\
&+ \frac{R^{\nu}_{0n}R^{\lambda}_{nm}R^{\mu}_{m0}}{(2\hbar\omega + \Omega_m + i\eta_m)(\hbar\omega + \Omega_n + i\eta_n)} \\
&+ \frac{R^{\lambda}_{0n}R^{\mu}_{nm}R^{\nu}_{m0}}{(\hbar\omega - \Omega_m + i\eta_m)(-\hbar\omega - \Omega_n - i\eta_n)} + (\lambda \leftrightarrow \upsilon)
\end{aligned} \tag{5}$$

Here, $\mu$, $\nu$, and $\lambda$ are Cartesian directions, $\epsilon_0$ is the vacuum permittivity, and $V$ is the volume of the crystal. The notation $(\lambda \leftrightarrow \upsilon)$ indicates an exchange of the two Cartesian directions. $R^{\upsilon}_{n0}$ are the optical transition matrix elements as discussed above. $R^{\nu}_{nm}$ are related to the optical coupling matrix elements between two exciton states, and the computation method of $R^{\nu}_{nm}$ can be found in Ref.[52]. $\eta_m$ and $\eta_n$ are the exciton-state-dependent broadening parameters. They are estimated by considering the quasiparticle lifetime due to electron-phonon scattering[65], using the approach described in Ref.[66].

Given that the experimentally measured SHG intensity follows $I_{\parallel}(\theta_\omega) \propto \left|\vec{E}(2\omega)\right|^2 \propto \left|\chi^{(2)}\right|^2 \left|\vec{E}(\omega)\right|^4$, we therefore qualitatively compare the calculated $|\chi^{(2)}|^2$ with the

experimental SHG spectra in Fig. 3. For the AA- and AB-stacked heterostructure, the SHG susceptibility tensor is not computed directly from the moiré supercell, as the supercell contains several hundreds of atoms, making BSE calculation computational prohibitive and ESC calculations even more challenging due to the required double summations over exciton states. Instead, the SHG susceptibility of the heterostructure is constructed by linearly combining the SHG susceptibilities of the two constituent monolayers. Specifically, for the AA stacking, the susceptibility is given by $\chi^{(2)}_{\mathrm{AA}} = \chi^{(2)}_{\mathrm{Mo}} + \chi^{(2)}_{\mathrm{W}}$, while for the AB stacking it is given by $\chi^{(2)}_{\mathrm{AB}} = \chi^{(2)}_{\mathrm{Mo}} - \chi^{(2)}_{\mathrm{W}}$, where a 180° in-plane rotation of the $WSe_2$ is assumed for the AB configuration. The above approximation is reasonable and can be understood as follows. As shown in Eq. (5), the SHG process involves three dipole-coupling matrix elements: the dipole couplings between the ground state and two intermediate excitons $|n\rangle$ and $|m\rangle$ (denoted by $R_{0n}$ and $R_{m0}$) , as well as the dipole coupling among these two intermediate excitons ($R_{nm}$). We note that the ground-to-exciton dipole matrix elements $R_{0n}$ are significant only for intralayer excitons, and that the inter-exciton coupling is large when both excitons originate from the same layer. As a result, the SHG response is dominated by contributions from intralayer excitons of the individual layers, which justifies the linear-combination approximation adopted here.

On the other hand, the intralayer exciton energies may be slightly modified by stacking. Based on our DFT calculation of the moiré supercell (see the next session), we estimate that the energy of the C exciton in $WSe_2$ is reduced by approximately 0.05-0.1 eV, while D excitons of $MoTe_2$ exhibit only a minor shift. Accordingly, in the calculation of $\chi^{(2)}_{\mathrm{AA}}$ and $\chi^{(2)}_{\mathrm{AB}}$, we apply a rigid redshift (0.08 eV) to $\chi^{(2)}_{\mathrm{W}}$ to account this effect.

**2. Estimation of stacking-induced energy shifts of intralayer excitons**

Large-scale DFT band-structure calculations are performed to estimate the stacking-induced energy shifts of intralayer excitons for both AA- and AB-stacked superlattices. The $MoTe_2/WSe_2$ superlattice is modeled using a 13×13 $MoTe_2$ supercell stacked on a 14×14 $WSe_2$ supercell, with a vacuum spacing greater than 20 Å. The Generalized gradient approximation (GGA) combined with DFT-D2 van der Waals correction[67] is employed, as implemented in the Vienna Ab Initio Simulation Package[68,69]. SOC is not included to reduce the computational cost, which is expected to have only a minor effect on the estimated excitonic energy shifts. The structure relaxation is performed with a force on each atom of less than 0.05 eV/Å. Owing to the large size of the moiré unit cell, gamma-point sampling (1×1×1) is employed for structure relaxation and self-consistent calculation. To compare with the original monolayer band structure, moiré band structures are unfolded onto $MoTe_2$ and $WSe_2$ layers using the VASPKIT post-processing package[70]. As shown in Supplemental Material Fig. S9[42], for both AA and AB stackings, near the M point, the bands unfolded onto the $MoTe_2$ layer closely resemble those of monolayer $MoTe_2$, indicating the energy of the D exciton of $MoTe_2$ remains nearly unchanged upon stacking. In contrast, for the $WSe_2$-projected bands, a clear reduction of the band gap is observed along the K-Γ path due to the stacking effect, suggesting a redshift of the C exciton energy. We estimate that the C-exciton energy is reduced by approximately 0.05-0.1 eV, and a value of 0.08 eV is adopted for the results shown in the main text.

## APPENDIX G: MODELLING THE SHG POLARIZATION

For monolayer TMDCs with $D_{3h}$ point group symmetry, the intrinsic second-order susceptibility tensor possesses four non-zero components: $\chi^{(2)}_{xxx} = -\chi^{(2)}_{yyx} = -\chi^{(2)}_{xyy} = -\chi^{(2)}_{yxy} = {\chi_0}^{(2)}$, where the $x$ and $y$ directions are defined along the AC and zigzag (ZZ) orientations of the TMDC, respectively. The SHG field can be written as $\vec{E}(2\omega) \propto {\chi_0}^{(2)}\vec{E}(\omega)\vec{E}(\omega)$, where $\vec{E}(\omega) = [E_x(\omega), E_y(\omega)]^T$. For a monolayer TMDC, the in-plane components of $\vec{E}(2\omega)$ are given by $\overrightarrow{E_x}(2\omega) \propto {\chi_0}^{(2)}(E_x(\omega)^2 - E_y(\omega)^2)$ and $\overrightarrow{E_y}(2\omega) \propto -2{\chi_0}^{(2)}E_x(\omega)E_y(\omega)$. Under normalized conditions, we can write $\vec{E}(\omega) = [cos\theta_\omega, sin\theta_\omega]^T$, where $\theta_\omega$ is the angle between $\vec{E}(\omega)$ and AC direction. Consequently, $\vec{E}(2\omega) = [cos2\theta_\omega, -sin2\theta_\omega]^T$. Hence, in a monolayer TMDC, a linearly polarized pump beam generates linearly polarized SHG, whose polarization azimuth is rotated by $-3\theta_\omega$ relative to the pump beam.

In the heterobilayer case, for simplicity, we set the bisector of the armchair directions of the two layers to be 0° (x direction, see the schematic diagram in Supplemental Material Fig. S10[42]). The two SHG field of the first and second TMDC monolayers can be expressed as:

$$\overrightarrow{E_1}(2\omega) = Ae^{i2\omega t}\begin{bmatrix}\cos(\frac{3\alpha}{2} - 2\theta_\omega) \\ \sin(\frac{3\alpha}{2} - 2\theta_\omega)\end{bmatrix} \tag{6}$$

$$\overrightarrow{E_2}(2\omega) = Be^{i(2\omega t - \Delta\phi)}\begin{bmatrix}\cos(\frac{3\alpha}{2} + 2\theta_\omega) \\ -\sin(\frac{3\alpha}{2} + 2\theta_\omega)\end{bmatrix} \tag{7}$$

where $A$ and $B$ is the amplitude of its SHG electric field of the first and second layer, $\Delta\phi$ is the phase difference between the SHG responses of the two layers, or equivalently, the phase difference between $\chi^{(2)}_{\mathrm{W}}$ and $\chi^{(2)}_{\mathrm{Mo}}$. Under the condition $A = B$ in simplification, the synthesized SHG field can be expressed as:

$$\vec{E}(2\omega) = A \begin{bmatrix} \cos\left(\frac{3\alpha}{2} - 2\theta_\omega\right) + e^{i(2\omega t - \Delta\phi)} \cos\left(\frac{3\alpha}{2} + 2\theta_\omega\right) \\ \sin\left(\frac{3\alpha}{2} - 2\theta_\omega\right) - e^{i(2\omega t - \Delta\phi)} \sin\left(\frac{3\alpha}{2} + 2\theta_\omega\right) \end{bmatrix} \quad (8)$$

The corresponding Stokes parameters are given by:

$$S_0 = |E_x(2\omega)|^2 + |E_y(2\omega)|^2 = 2A^2(1 + \cos 3\alpha \cos\Delta\phi) \quad (9)$$

$$S_1 = |E_x(2\omega)|^2 - |E_y(2\omega)|^2 = 2A^2(\cos 3\alpha + \cos\Delta\phi) \cos 4\theta_\omega \quad (10)$$

$$S_2 = 2Re\left[E_x(2\omega)E_y^*(2\omega)\right] = -2A^2(\cos 3\alpha + \cos\Delta\phi) \sin 4\theta_\omega \quad (11)$$

$$S_3 = 2Im\left[E_x(2\omega)E_y^*(2\omega)\right] = -2A^2 \sin 3\alpha \sin \Delta\phi \quad (12)$$

, satisfying $S_0^2 = S_1^2 + S_2^2 + S_3^2$ with linear polarization state. The ellipticity angle $\chi$ can be obtained by $\chi = \frac{1}{2}\arcsin(\frac{S_3}{S_0})$ and ellipticity is $\tan \chi$, both of which become independent of $\theta_\omega$. The polarization ellipticity can also be defined as the ratio between the semi-minor and semi-major axes of the polarization ellipse. The azimuth angle $\theta_{2\omega}$ of the polarization ellipse satisfies:

$$\tan(2\theta_{2\omega}) = \frac{S_2}{S_1} = -\tan 4\theta_\omega \quad (13)$$

$$\theta_{2\omega} = -2\theta_\omega + \begin{cases} 0, & \Delta\phi < 180° - 3\alpha \\ \frac{\pi}{2}, & \Delta\phi > 180° - 3\alpha \end{cases} \quad (14)$$

When $\cos 3\alpha + \cos\Delta\phi = 0$ or equivalently $\Delta\phi + 3\alpha = 180°$, the SHG output becomes perfectly circularly polarized, with the handedness determined by the sign of $S_3$. That is to say, regardless of the value of $\theta_\omega$, when the sign of $\cos 3\alpha + \cos\Delta\phi$ changes, $\theta_{2\omega}$ will change by 90°. Given that in our examples $\sin 3\alpha \sin \Delta\phi > 0$, we have $S_3 < 0$, corresponding to left-handed circular polarization. Hence, when $\Delta\phi = 180° - 3\alpha$, we get a left-handed circular SHG output.

The length of the semi-major and semi-minor axis of the polarization ellipse can be written as:

$$\sqrt{I_{\max,\min}} = \sqrt{\frac{S_0 \pm \sqrt{{S_1}^2 + {S_2}^2}}{2}} = A\sqrt{1 + \cos 3\alpha \cos\Delta\phi \pm |\cos 3\alpha + \cos\Delta\phi|} \quad (15)$$

Notably, when $\Delta\phi = 180° - 3\alpha$, the semi-major (semi-minor) axis reaches a minimum (maximum), as shown in Supplemental Material Fig. S7[42]. To plot Poincaré-sphere, we use normalized Stokes parameters with $s_i = S_i/S_0$ (*i*= 1, 2, 3).

In the parallel configuration used in our experiments, the SHG signal is projected onto the polarization direction of the incident light. The total polarization-dependent SHG intensity after coherent superposition is then given by:

$$I_{\|}(\theta_{\omega}) \propto \left|\left(\overrightarrow{E_1}(2\omega)+\overrightarrow{E_2}(2\omega)\right)\cdot\overrightarrow{e_E}\right|^2 \quad (16)$$

$$= \left|Ae^{i2\omega t}\cos 3\left(\frac{\alpha}{2}-\theta_{\omega}\right)+Be^{i(2\omega t-\Delta\phi)}\cos 3\left(\frac{\alpha}{2}+\theta_{\omega}\right)\right|^2$$

$$= A^2\cos^2 3(\tfrac{\alpha}{2}-\theta_{\omega})+B^2\cos^2 3(\tfrac{\alpha}{2}+\theta_{\omega})+2AB\cos 3(\tfrac{\alpha}{2}-\theta_{\omega})\cos 3(\tfrac{\alpha}{2}+\theta_{\omega})\cos\Delta\phi$$

where $\overrightarrow{e_E}$ is the unit vector in the direction of $\vec{E}(\omega)$. This expression serves as the model for the simulations presented in Fig. 4(d) and Supplemental Material Fig. S7[42]. In these simulations, we set $A$ = $B$ = 210 (representative of the SHG amplitude of monolayer $WSe_2$, which is similar to that of monolayer $MoTe_2$ within the experimental range).

Under the condition $A$ = $B$, Eq. (8) could be simplified to:

$$I_{\|}(\theta_{\omega}) \propto A^2[1+\cos\Delta\phi\cos 3\alpha+\cos 6\theta_{\omega}(\cos\Delta\phi+\cos 3\alpha)] \quad (17)$$

From the third term in equation (9), it can be seen that when the condition $\cos\Delta\phi+\cos 3\alpha=0$ (i.e., $\Delta\phi$ = 180° − 3α) is satisfied, $I_{\|}$ becomes independent of $\theta_{\omega}$, and circularly polarized SHG is emitted. When $\Delta\phi < 180° - 3\alpha$, $I_{\|}$ reaches its maximum value $I_{max}$ at $\theta_{\omega}$ = 0°, whereas when $\Delta\phi > 180° - 3\alpha$, $I_{\|}$ reaches $I_{max}$ at $\theta_{\omega}$ = 30°. Moreover, since $I_{\|}$ reaches its maximum $I_{max}$ only when the third term in Eq. (9) is positive, it follows that at $\Delta\phi = 180° - 3\alpha$, where this term vanishes, $I_{max}$ attains a minimum in the parameter space, corresponding to SHG destructive interference.

## APPENDIX H: TUNABLE CLASSICAL AND QUANTUM LIGHT SOURCES WITH TMDC HETEROBILAYERS

Based on the principles proposed here, TMDC heterobilayers offer promising opportunities for developing tunable classical and quantum light sources. One approach involves the realization of a frequency-doubling source with tunable polarization, where the polarization state can be continuously adjusted via changes in $\Delta\phi$ and $\alpha$, and is uniquely linked to the excitation energy. Another possibility lies in spontaneous parametric down-conversion[71-73], the reverse process of SHG, where a single $2\omega$ photon is converted into two coherent $\omega$ photons, with the capability to tailor the entanglement entropy of the photon pair. In Ref.[74], tunable SHG polarization is achieved in monolayer $MoS_2$ by adjusting the delay between two perpendicular pulses; however, this method requires a complex external optical setup. Ref.[75] realized similar functionality by varying the interlayer spacing and twist angle between two $h$-BN flakes. In contrast, TMD heterobilayers offer significant advantages. Compared to the microelectromechanical twisted $h$-BN system, TMDC heterobilayers offer distinct advantages, including a significantly larger second-order nonlinear susceptibility $\chi^{(2)}$ (tens of times greater than that of $h$-BN) and greatly reduced fabrication complexity.

According to the derived expression for circularly polarized intensity: $I_{\|} = A^2\sin^2 3\alpha$, the maximum output reaches $A^2$ (the single-layer SHG intensity) when $\alpha$ = 30° and $\Delta\phi$ = 90° (Supplemental Material, Fig. S7[42]). For TMDC, the excitation energy yielding the highest SHG efficiency corresponds to the C/2 exciton resonance (as demonstrated in Section 3). Therefore, the energy at which $\Delta\phi$ = 90° should lie near the C/2 exciton energies of both materials (around 1.3-1.5 eV). Hence, our findings also offer a new perspective for developing highly compact, reconfigurable, and robust nonlinear and quantum optical devices based on TMDC heterobilayers.

## Supplementary Figures

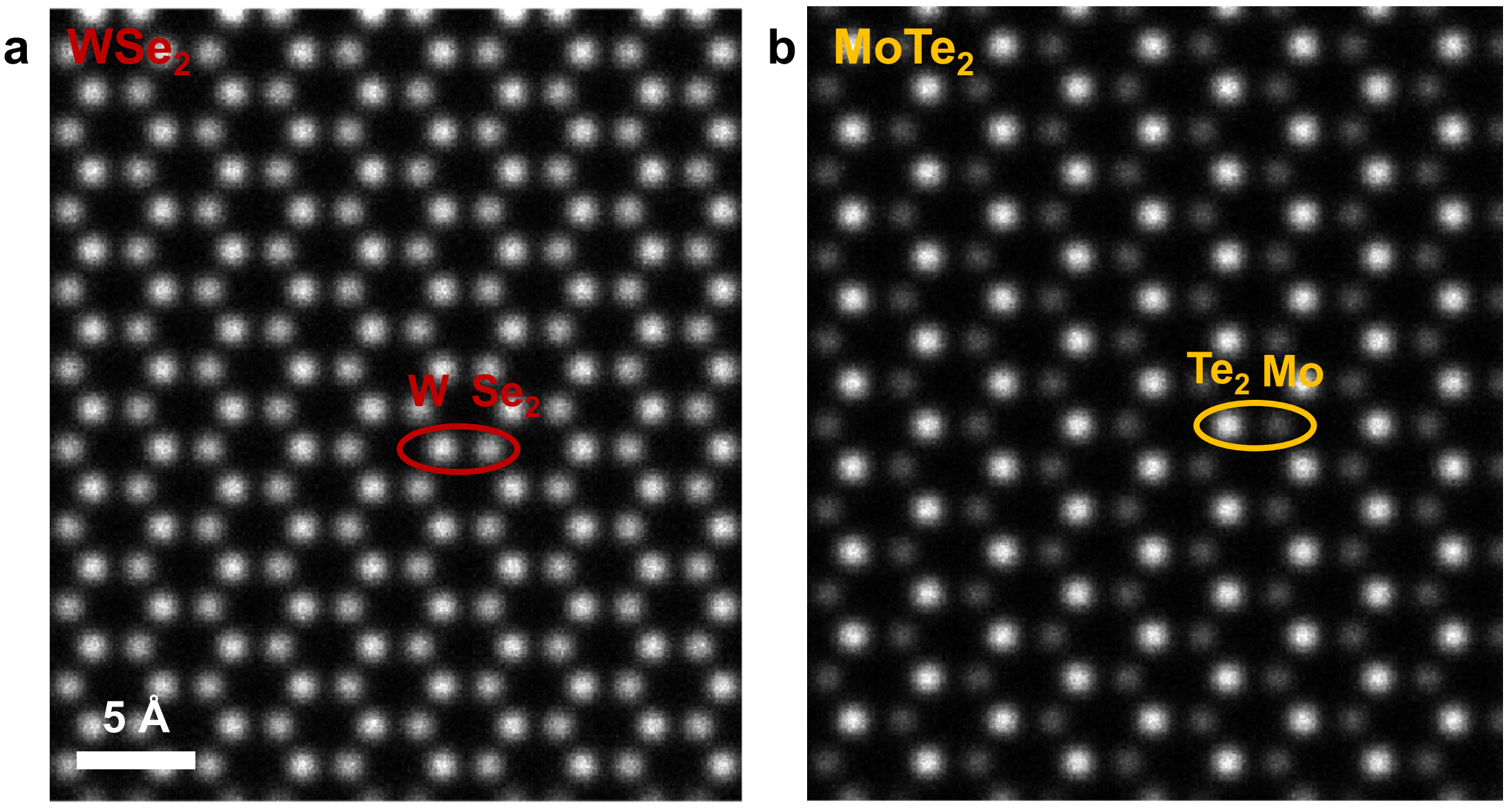

**Supplementary Figure S1 | a-b,** Simulated ADF images (collection angle 54-200 mrad) of single-layer (a) $WSe_2$ and (b) $MoTe_2$.

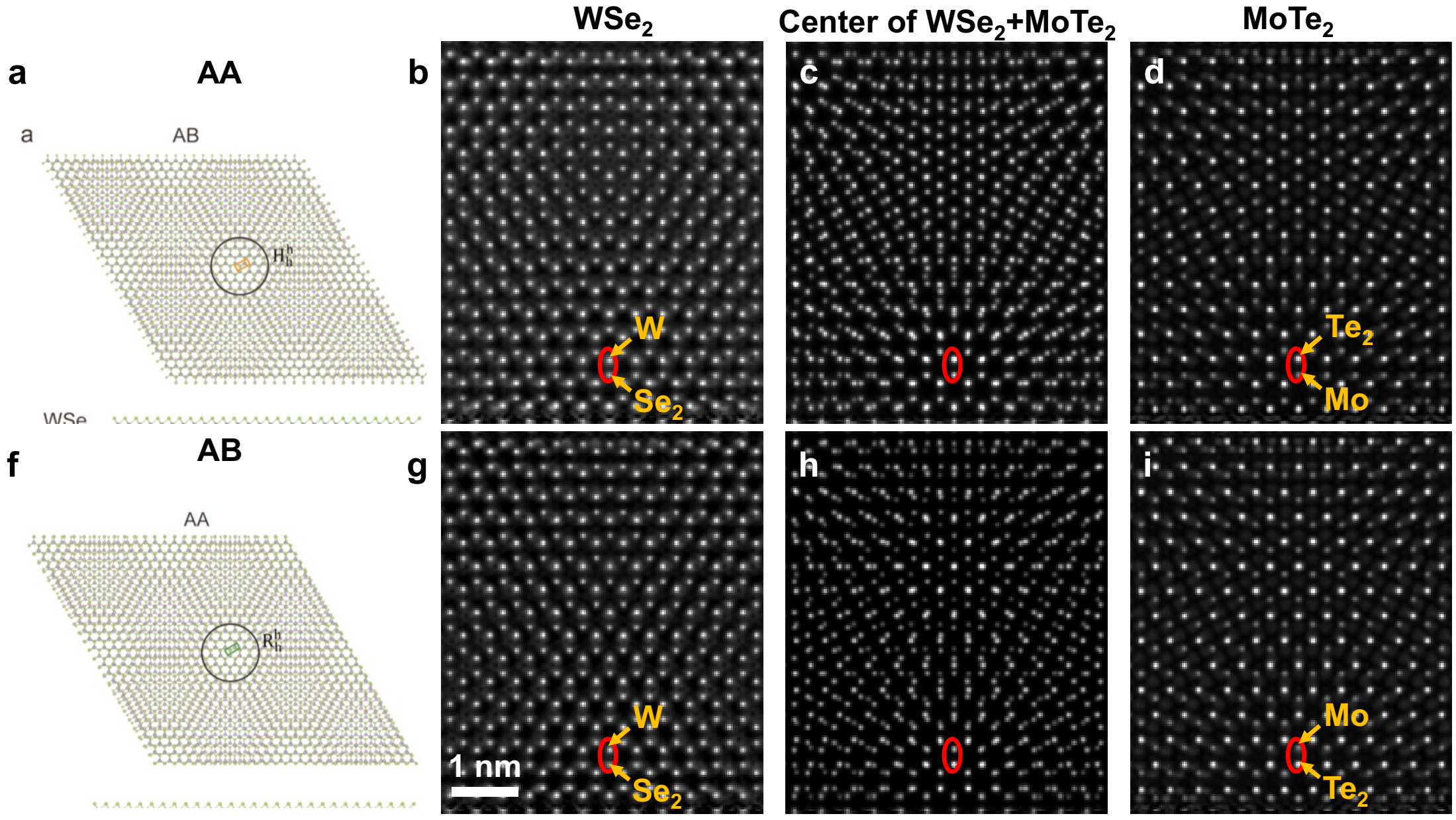

**Supplementary Figure S2 | a,f,** Atomic model of **(a)** AA and **(f)** AB stacking of $MoTe_2/WSe_2$. **b-d,** Simulated MEP phase images at different depth of AA stacking of $MoTe_2/WSe_2$. **g-i,** Simulated MEP phase images at different depth of AB stacking of $MoTe_2/WSe_2$.

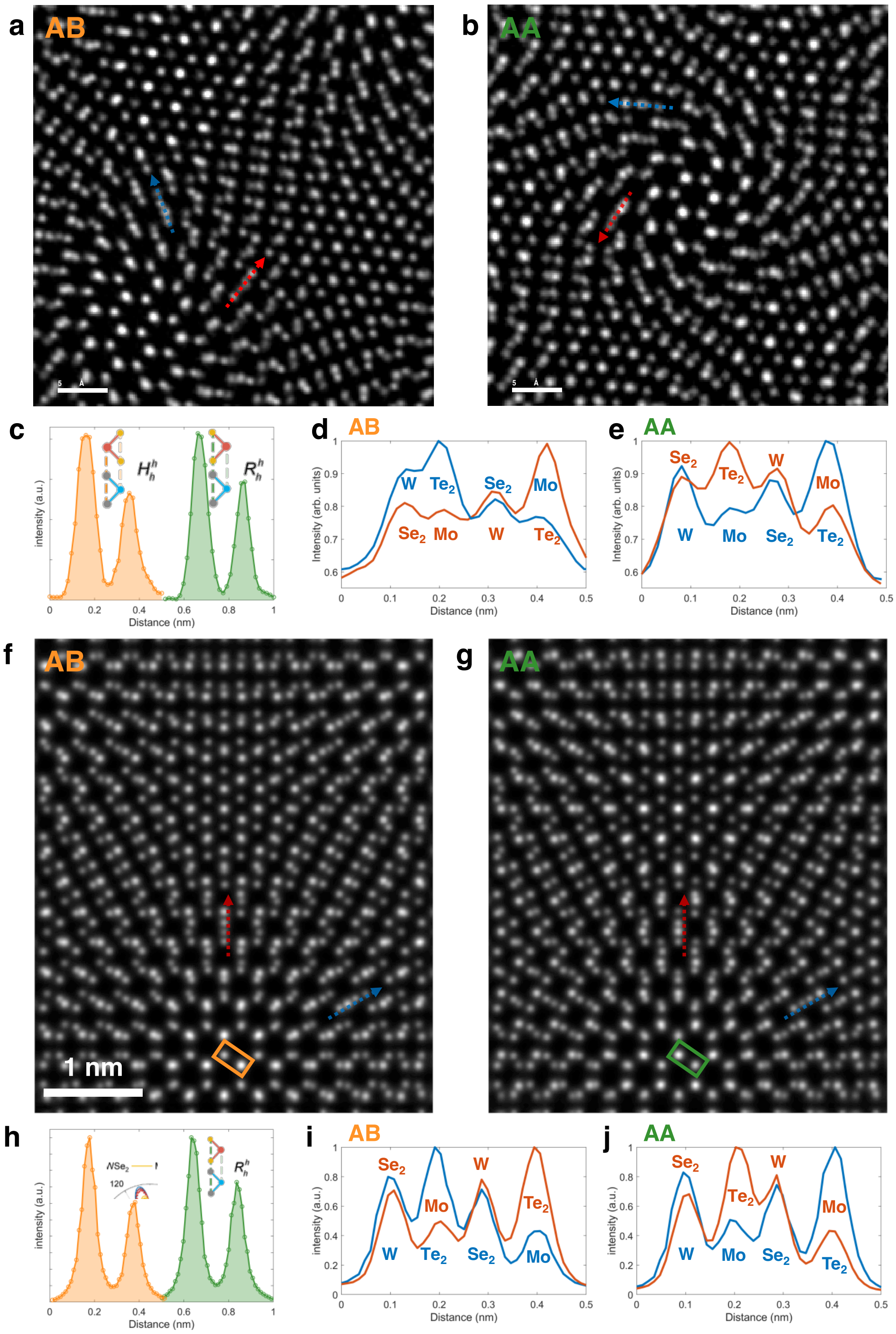

**Supplementary Figure S3 | Comparison of experimental and simulation results. a,b,** Experimental MEP phase images of **(a)**AB and **(b)**AA stacking of $MoTe_2/WSe_2$. **c,** Intensity profile across the orange and green frame in **Fig. 1i** and **Fig. 1j**, corresponding to $R_h^h$ or $H_h^h$ region. **d,e,** intensity profiles corresponding the labelled lines. **f,g,** Simulated MEP phase image of AB and AA stacking of $MoTe_2/WSe_2$. **h,** Intensity profiles across the orange and green region in **(f)** and **(g)**, corresponding to $R_h^h$ or $H_h^h$ region. **i,j,** Intensity profiles across the four kinds of atom columns from the labelled lines in **(f), (g)**.

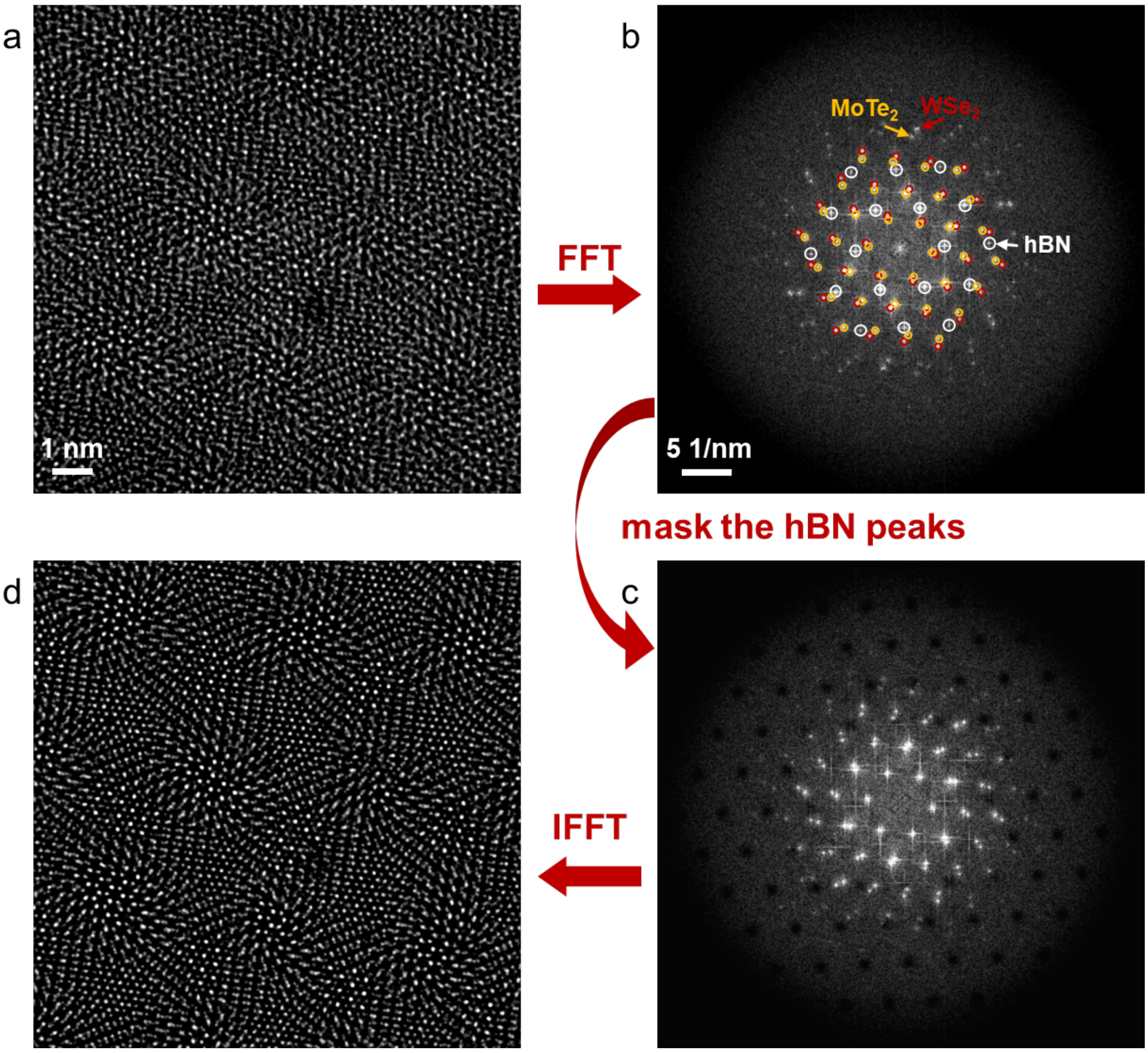

**Supplementary Figure S4 | Image processing to remove the structural mixture of encapsulated hBN. a**, MEP image of the $MoTe_2/WSe_2$ bilayer with the structural mixture of encapsulated hBN. **b**, fast Fourier transformation of **(a)**, illustrating the presence of $MoTe_2$, $WSe_2$, and hBN. **c**, Fourier transformation masking the hBN peaks with an 8-pixel edge smooth. **d**, Inverse fast Fourier transform of **(c)**, displaying the enhanced contrast of the $MoTe_2/WSe_2$ bilayer.

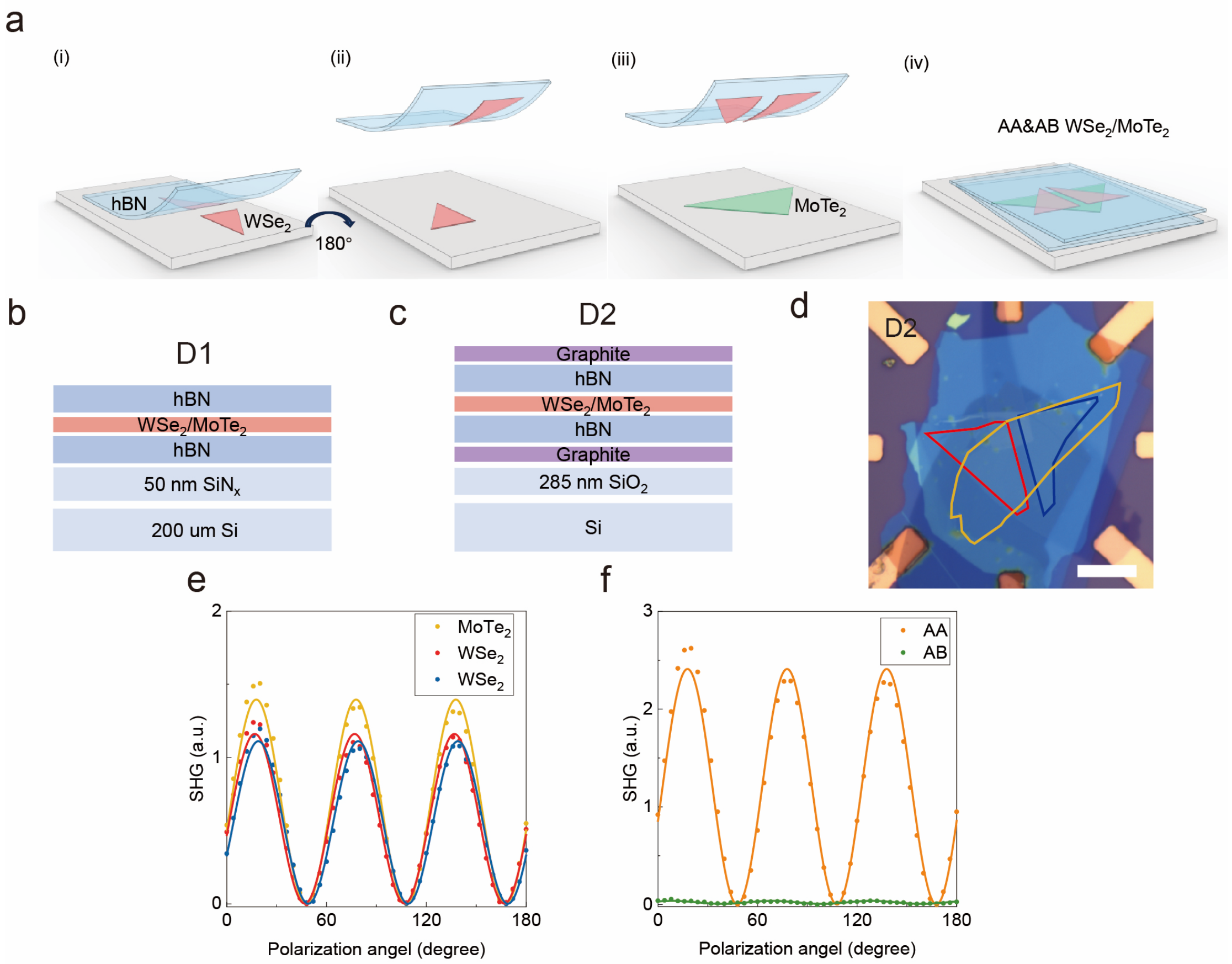

**Supplementary Figure S5 | Device structures and characterization of heterostructures D1 and D2. a,** Schematic procedures for deterministically creating both AA and AB stacked TMDC moiré heterobilayers within a single device. **b, c,** Schematic diagrams of devices D1 and D2. **d,** Optical micrograph of device D2. Red and blue outlines indicate two $WSe_2$ flakes rotated 180° relative to each other; yellow indicates the $MoTe_2$ flake. **e, f,** Angle-resolved SHG of three monolayer regions and two heterostructure regions in device D2.

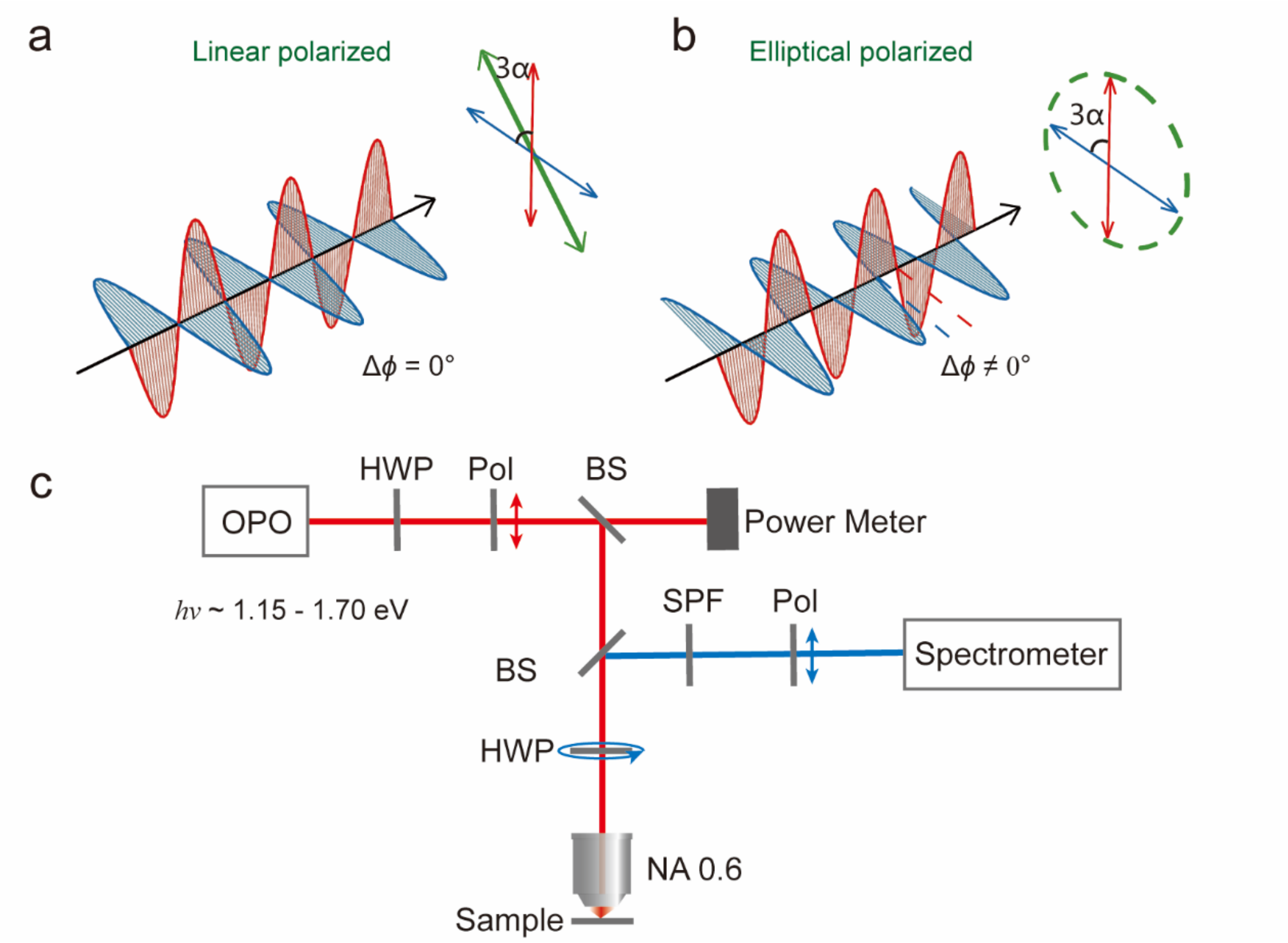

**Supplementary Figure S6 | Schematic diagrams of superposition of two linearly polarization states. a,** Two linearly polarization states with $\Delta\phi = 0$. **b,** Two linearly polarization states with $\Delta\phi \neq 0$. **c,** A sketch of the set-up used for the experiments. We used an optical parametric oscillator (OPO) for tunable excitation energy between ~1.15 and 1.70 eV. HWP, half-wave plate; Pol, polarizer; BS, beam splitter; SPF, short pass filter.

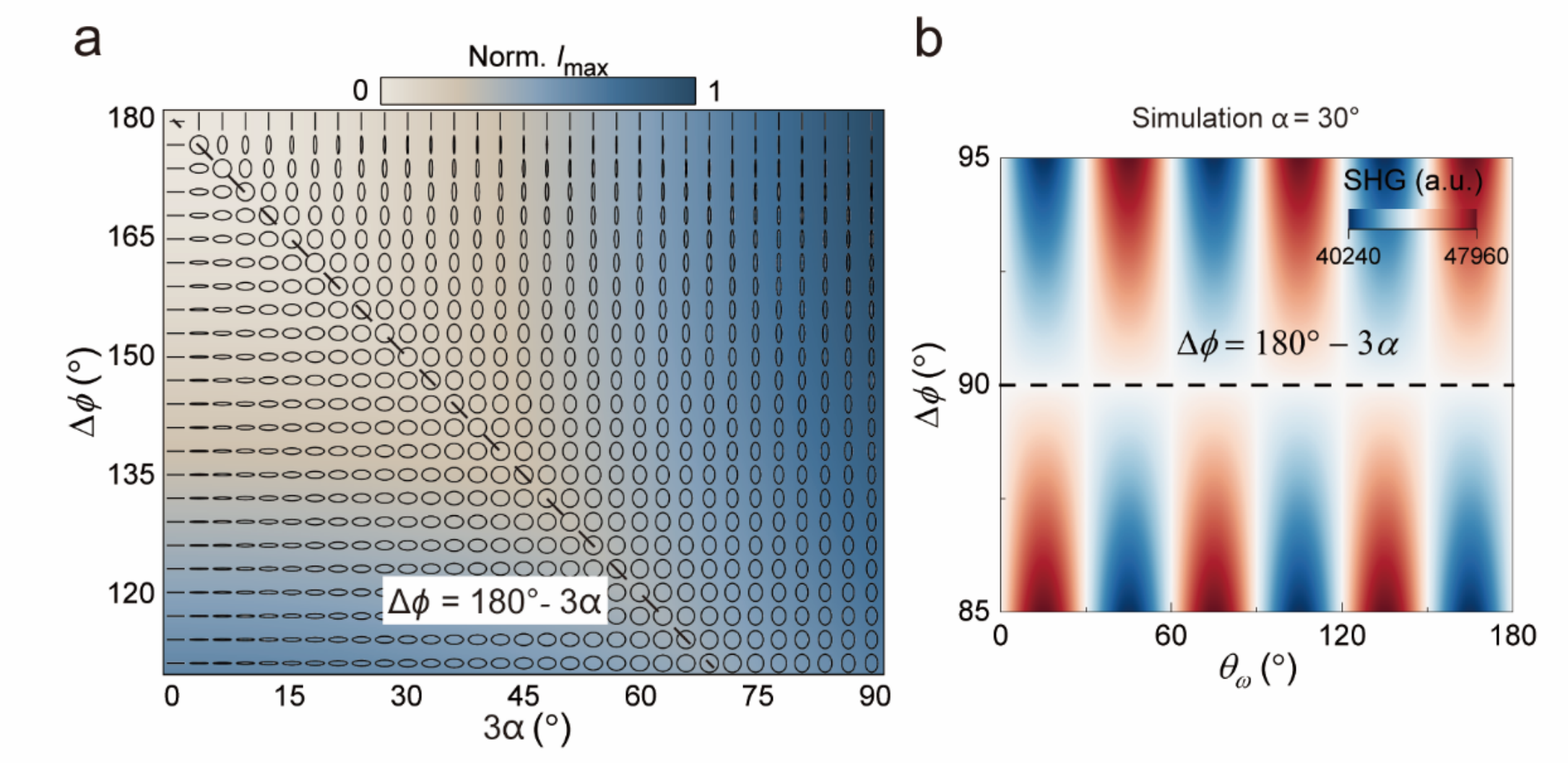

**Supplementary Figure S7 | Additional simulated polarization-dependent SHG intensity. a,** Simulation of normalized $I_{max}$ as a function of $\Delta\phi$ and $\alpha$, with each ellipse representing a polarization state. The dashed line corresponds to a minimum of $I_{max}$, which is also the condition for circularly polarized SHG. **b,** Simulated $\Delta\phi$ and polarization-dependent SHG intensity for $\alpha = 30°$. Circularly polarized SHG is emitted at $\Delta\phi = 180° - 3\alpha$. The maximum SHG intensity for $\alpha = 90°$ is two orders of magnitude larger than that for $\alpha = 1°$ (Fig. 4f), highlighting the critical role of designed $\alpha$ and $\Delta\phi$.

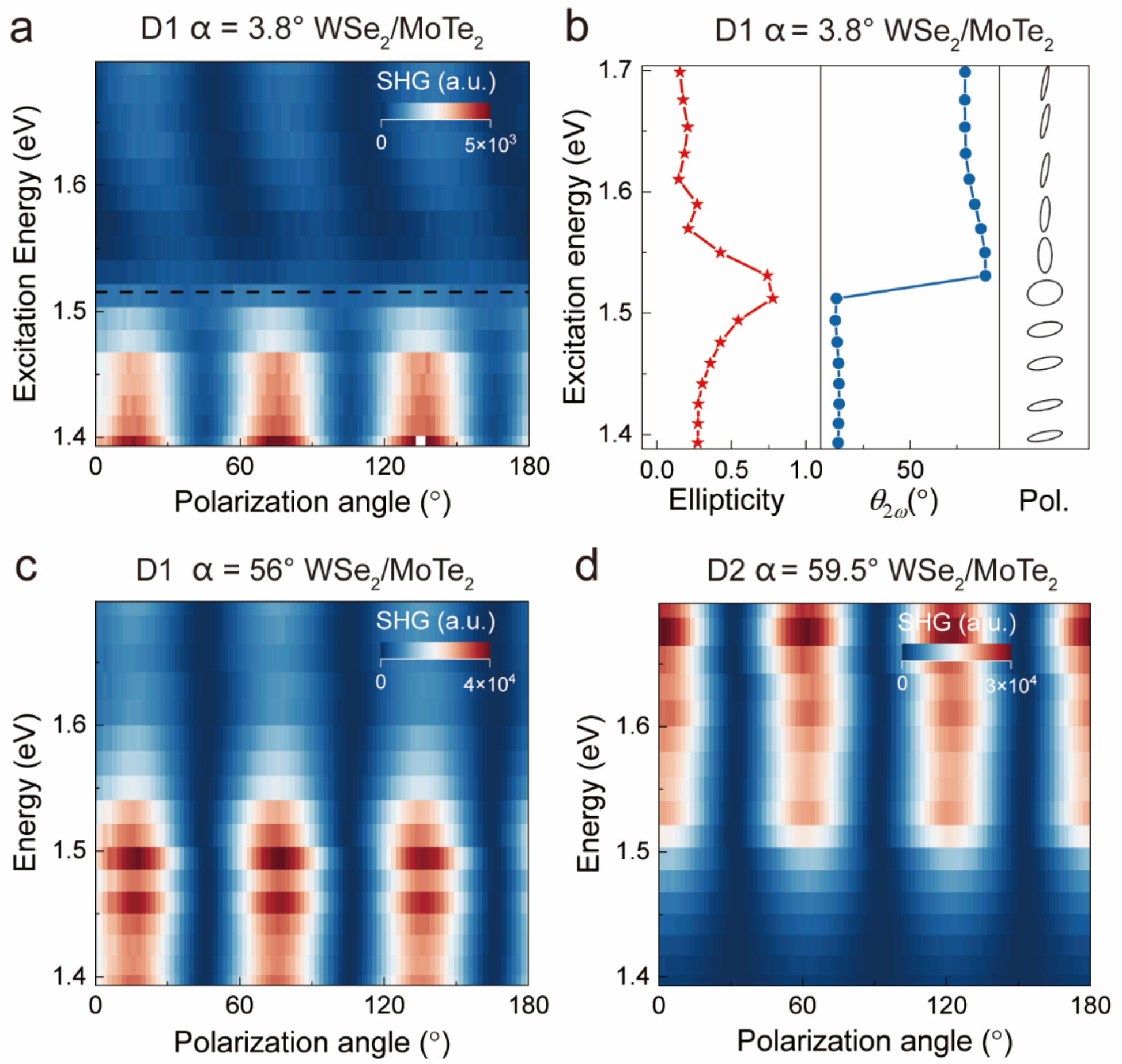

**Supplementary Figure S8 | Additional measured energy- and polarization-dependent SHG intensity. a, c, d, e, f,** SHG as a function of excitation-photon-energy and polarization, measured from **a,** D1 ($\alpha$ = 3.8° region), **c,** D1 ($\alpha$ = 56° region), **d,** D2 ($\alpha$ = 59.5° region). For all devices, there are no significant polarization change in near-AB-stacked regions. **b,** SHG ellipticity, azimuthal angle $\theta_{2\omega}$ and polarization states as functions of excitation photon energy, extracted from **(a).**

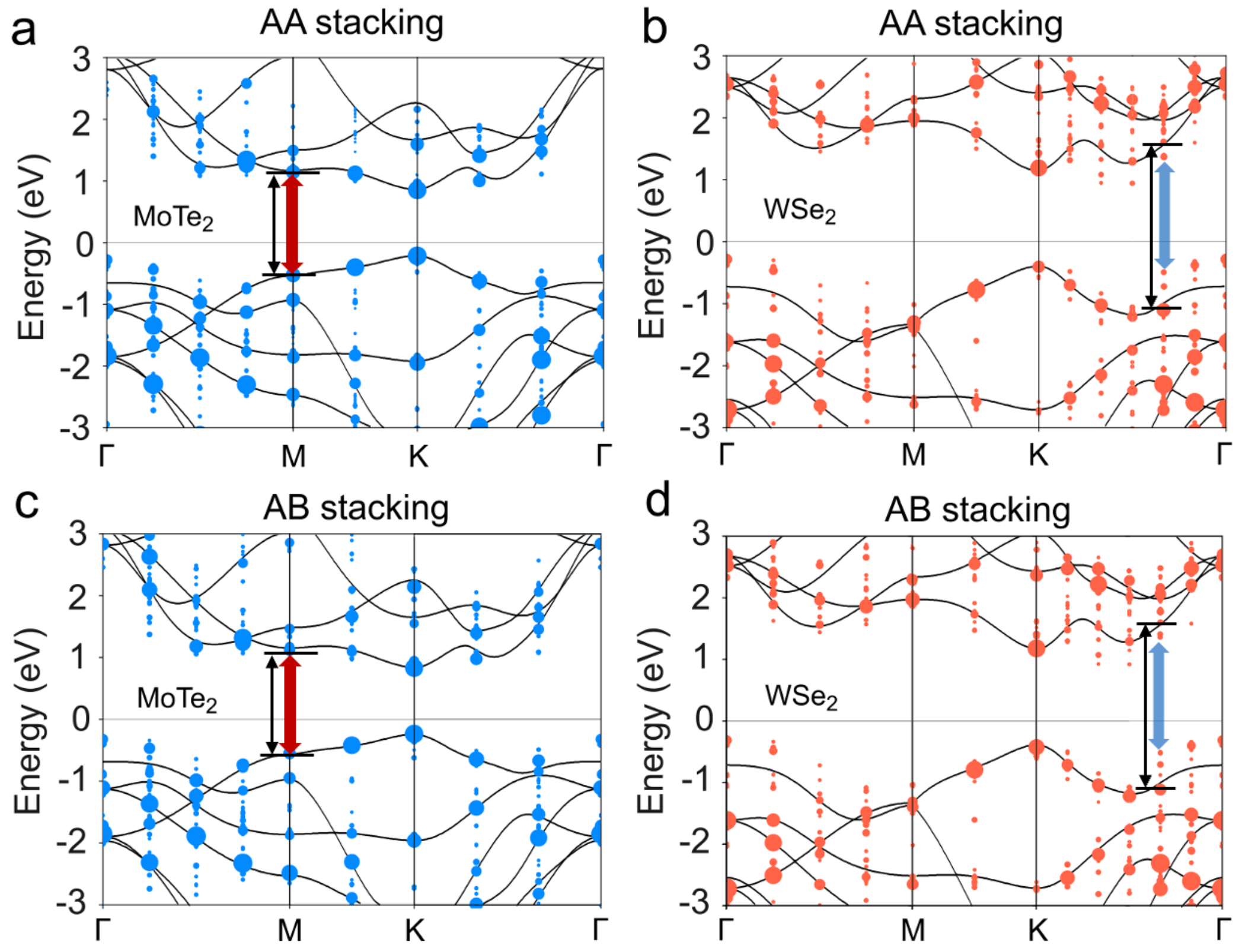

**Supplementary Figure S9 | Unfolded band structures for $MoTe_2/WSe_2$ superlattice. a, b,** Unfolded band structures of AA-stacked $WSe_2/MoTe_2$ projected onto (**a**) the $MoTe_2$ layer and its primitive BZ (blue dots) and (**b**) the $WSe_2$ layer and its primitive BZ (red dots). The size of the dots denotes the weight of the unfolding. Black solid lines are the band structures of the corresponding monolayers. **c,d,** Similar to (**a**) and (**b**), but for AB-stacked $WSe_2/MoTe_2$. Notably, for both configurations, the stacking effect negligibly affects the optical transition energy associated with D exciton of $MoTe_2$ layer (indicated by the red double-headed arrows), while effectively reducing that associated with C excitons in $WSe_2$ (indicated by the blue double-headed arrows). SOC is not included in the above calculations, which would not affect our conclusion.

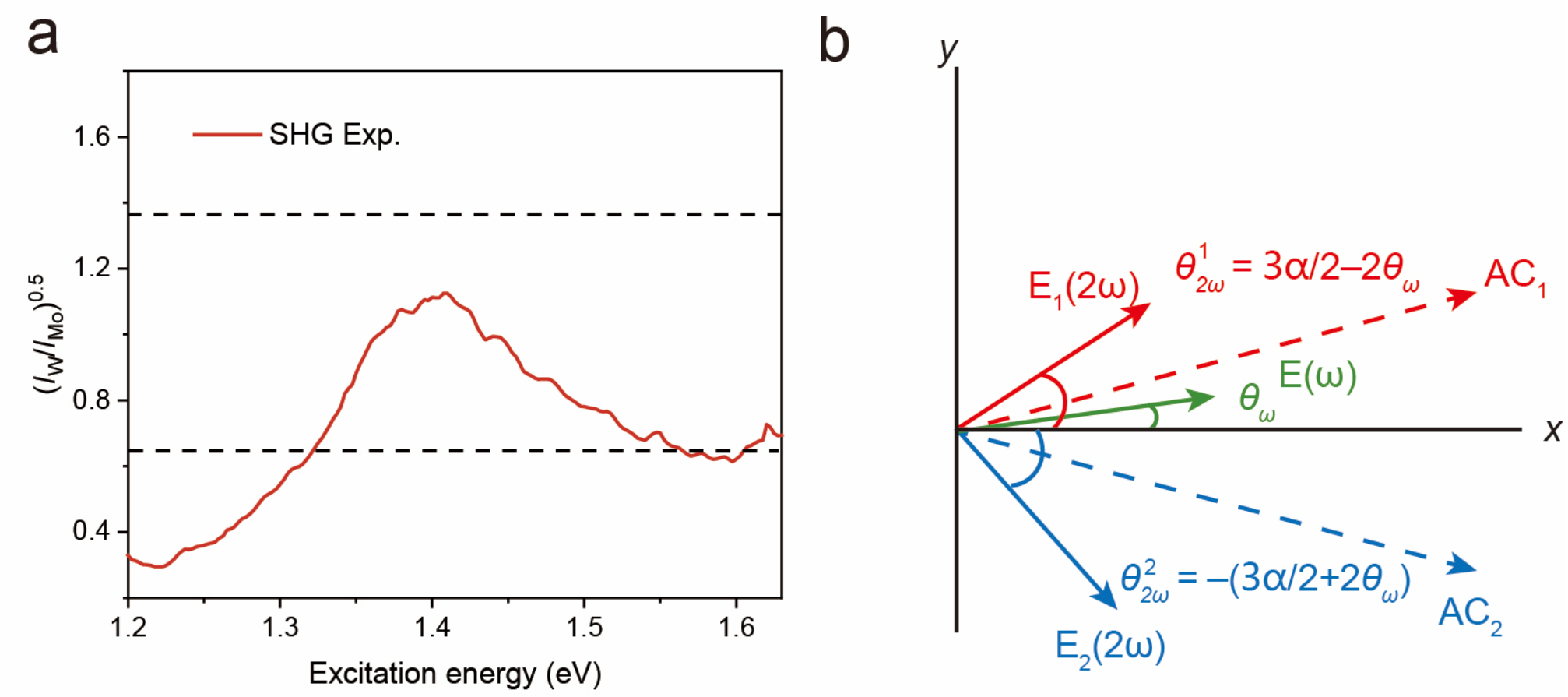

**Supplementary Figure S10 | a,** Calculated $A/B \propto \sqrt{I_W/I_{Mo}}$ from the experimental SHG intensity of $WSe_2$ $I_W$ and $MoTe_2$ $I_{Mo}$, within the excitation energy of 1.321 to 1.565 eV, $A/B \in (0.65,1.35)$. Here, we apply a rigid redshift (0.08 eV) to the SHG spectrum of $WSe_2$ to account the stacking effect, same as first principles calculations. **b,** Schematics considering two layers with the armchair ($AC_1$ and $AC_2$, indicated by dashed red and blue arrows, respectively) orientation misalignment of $\alpha$. The SHG polarization angles are $\theta^1_{2\omega} = 3\alpha/2-2\theta_\omega$ for the first layer (red solid arrow) and $\theta^2_{2\omega} = -(3\alpha/2+2\theta_\omega)$ for the second layer (blue solid arrow).